\documentclass[sigconf]{acmart}

\AtBeginDocument{%
  \providecommand\BibTeX{{%
    \normalfont B\kern-0.5em{\scshape i\kern-0.25em b}\kern-0.8em\TeX}}}

\makeatletter
\def\@ACM@checkaffil{
    \if@ACM@instpresent\else
    \ClassWarningNoLine{\@classname}{No institution present for an affiliation}%
    \fi
    \if@ACM@citypresent\else
    \ClassWarningNoLine{\@classname}{No city present for an affiliation}%
    \fi
    \if@ACM@countrypresent\else
        \ClassWarningNoLine{\@classname}{No country present for an affiliation}%
    \fi
}

\copyrightyear{2024}
\acmYear{2024}
\setcopyright{acmlicensed}
\acmConference[KDD '24] {Proceedings of the 30th ACM SIGKDD Conference on Knowledge Discovery and Data Mining }{August 25--29, 2024}{Barcelona, Spain.}
\acmBooktitle{Proceedings of the 30th ACM SIGKDD Conference on Knowledge Discovery and Data Mining (KDD '24), August 25--29, 2024, Barcelona, Spain}
\acmISBN{979-8-4007-0490-1/24/08}
\acmDOI{10.1145/3637528.3671520}

\usepackage[super]{nth}
\usepackage{amsmath}
\usepackage{amsthm}
\usepackage{mathtools}
\usepackage{bm}
\usepackage{dsfont}
\usepackage{graphicx}
\usepackage{caption}
\usepackage{subcaption}
\usepackage{float}
\usepackage{enumitem}
\usepackage{array}
\usepackage{makecell}
\usepackage{multirow}
\usepackage{tabularx}
\usepackage{url}
\usepackage{xcolor}
\usepackage[flushleft]{threeparttable}
\usepackage[ruled,vlined,linesnumbered]{algorithm2e}

\theoremstyle{definition}

\begin{document}

\title{Unified Low-rank Compression Framework for Click-through Rate Prediction}

\author{Hao Yu}
\email{yuh@lamda.nju.edu.cn}
\affiliation{%
  \institution{Nanjing University}
  \city{Nanjing}
  \state{Jiangsu}
  \country{China}
}

\author{Minghao Fu}
\email{fumh@lamda.nju.edu.cn}
\affiliation{%
  \institution{Nanjing University}
  \city{Nanjing}
  \state{Jiangsu}
  \country{China}
}

\author{Jiandong Ding}
\email{jdding.cs@gmail.com}
\affiliation{%
   \institution{Researcher}
   \city{Shanghai}
   \country{China}
}

\author{Yusheng Zhou}
\email{zhou-yusheng@foxmail.com}
\affiliation{%
\institution{Researcher}
\city{Shanghai}
\country{China}
}

\author{Jianxin Wu}
\email{wujx2001@nju.edu.cn}
\affiliation{%
  \institution{Nanjing University}
  \city{Nanjing}
  \state{Jiangsu}
  \country{China}
}

\renewcommand{\shortauthors}{Hao Yu, Minghao Fu, Jiandong Ding, Yusheng Zhou \& Jianxin Wu}

\begin{abstract}
  Deep Click-Through Rate (CTR) prediction models play an important role in modern industrial recommendation scenarios. However, high memory overhead and computational costs limit their deployment in resource-constrained environments. Low-rank approximation is an effective method for computer vision and natural language processing models, but its application in compressing CTR prediction models has been less explored. Due to the limited memory and computing resources, compression of CTR prediction models often confronts three fundamental challenges, i.e., (1). How to reduce the model sizes to adapt to edge devices? (2). How to speed up CTR prediction model inference? (3). How to retain the capabilities of original models after compression? Previous low-rank compression research mostly uses tensor decomposition, which can achieve a high parameter compression ratio, but brings in AUC degradation and additional computing overhead. To address these challenges, we propose a unified low-rank decomposition framework for compressing CTR prediction models. We find that even with the most classic matrix decomposition SVD method, our framework can achieve better performance than the original model. To further improve the effectiveness of our framework, we locally compress the output features instead of compressing the model weights. Our unified low-rank compression framework can be applied to embedding tables and MLP layers in various CTR prediction models. Extensive experiments on two academic datasets and one real industrial benchmark demonstrate that, with 3-5$\times$ model size reduction, our compressed models can achieve both faster inference and higher AUC than the uncompressed original models. Our code is at \href{https://github.com/yuhao318/Atomic_Feature_Mimicking}{https://github.com/yuhao318/Atomic\_Feature\_Mimicking}.
\end{abstract}



\maketitle
\section{Introduction}\label{sec-introduction}
 Benefiting from the large model sizes and powerful neural architectures, deep learning recommendation models (DLRMs) show unparalleled advantages in mining users' potential interests. To improve the capabilities of such models, a common method in practice is to increase the model sizes. However, these recommender systems rely heavily on abundant storage, huge memory access, and large computing resources for frequent and swift inference requested by millions of concurrent users. Although enlarging the model size is effective, it can also become a major obstacle to model deployment and real-time predictions, especially for devices with limited resources. Therefore, recommendation model compression has attracted immense interest in recent years.

Generally speaking, modern deep CTR prediction models can be divided into three main modules, i.e., the embedding tables process categorical features by encoding sparse, high-dimensional inputs into a dense vector representation (i.e. embeddings), the MLP layers are used for feature interaction modeling and prediction, and the feature interaction module learns cross representations of high-order features. Each feature field has a unique embedding, which is stored in the embedding tables. However, in actual industrial scenarios, there are usually billions or even trillions of categorical features, and the embedding tables may require hundreds of GB or even TB storage space. As a result, the embedding layers occupy most of the parameters in a CTR prediction model. Moreover, the MLP layers correspondingly occupy most of the computational cost during the inference process, but the effort to compress it remains scarce. Therefore, we believe that to efficiently and economically deploy a CTR prediction model in an actual production system, its embedding tables and MLP layers both need to be compressed.

Low-rank decomposition methods, including matrix decomposition and tensor decomposition methods, are widely used in various model compression tasks. The most classic method of matrix decomposition is Singular Value Decomposition (SVD). However, it is rarely explored in the field of compressing deep CTR prediction models. Traditional tensor decomposition methods, such as Tensor-Train Decomposition (TTD), need to re-multiply all tensors to obtain the embedding during inference, so it will introduce abundant additional computational overhead. However, the embedded tables of industrial DLRMs may reach hundreds of GB or even TB levels, far exceeding the maximum tens of GB memory capacity of a single GPU, so most companies deploy their recommendation models on CPUs. Therefore, the decompression calculation caused by tensor decomposition will greatly slow down the model's inference speed. Besides, tensor decomposition is also a lossy compression method that reduces the models' AUC, and every 1\textperthousand ~drop in AUC represents a significant loss of economic benefits. Moreover, Tensor-Train Decomposition is only applicable to the embedding layer, and there is currently a lack of research on applying tensor decomposition in the MLP layers.

Hence, we believe that in order to successfully perform low-rank approximation to compressing CTR prediction models, we need an effective framework that can be easily generalized into various DLRMs, reduces the embedding layer sizes to achieve a higher parameter compression ratio, further compresses the MLP layers to obtain faster model acceleration, and still maintains or even improves the models' capacity after compression.

To fulfill these goals, we propose a unified low-rank decomposition method, which can be used on both the MLP layers and embedding tables of the CTR prediction models. With our compression framework, even the classical matrix decomposition SVD method can obtain better performances than the original models. To further improve the AUC of the compressed models, we design a novel way to replace SVD with Atomic Feature Mimicking (AFM) ~\cite{yu2020compressing}. AFM's idea stems from a simple but crucial realization: when compressing a deep learning model, we should focus on minimizing the loss of the model outputs, rather than the weight~\cite{luo2017thinet,yu2020compressing}. Therefore, AFM utilizes the PCA technology to low-rank approximate MLP outputs. Inspired by this idea, to compress the MLP layers, we introduce and improve AFM to make it work properly for MLP in CTR prediction models. To further improve the model's performance, we add an extra activation function between the two decomposed linear layers. Then, for the large-scale embedding layers of the CTR prediction models, we further extend AFM for the embedding tables to reduce its dimensions. The compressed weights can be incorporated into the original model to achieve further acceleration. Since these two compressing methods are orthogonal, we can naturally combine them to obtain higher compression ratios. Our contributions are as follows: 

\begin{itemize}
    \item Our paper analyzes issues with traditional low-rank decomposition, i.e., its limited scope of application and poor performance. Therefore, we propose a unified low-rank decomposition framework for compressing the MLP layers and the embedding tables of the CTR prediction models.
    \item Unlike standard low-rank decomposition, our plug-and-play framework can be easily generalized to mainstream CTR prediction models and greatly improve their AUC by mimicking feature distributions to achieve better output approximation.
    \item Abundant experiments prove the effectiveness of our framework. On two academic datasets and a real industry recommendation dataset, our methods achieve 3-5$\times$ compression ratios, significantly outperform the original models' AUC scores, and largely reduce the models' inference time.
\end{itemize}

\section{Related Work}\label{sec-relatedwork}

Our work is connected to several themes in the literature, which we describe next.

\subsection{Deep CTR Prediction Models}

Many deep learning recommendation models have been proposed over the past years and have achieved state-of-the-art performances in CTR prediction tasks. Wide \& Deep~\cite{cheng2016wide} jointly trains wide linear models and deep neural networks to combine the benefits of memorization and generalization for recommender systems. DCN~\cite{wang2017deep} introduces a novel cross-network to capture certain bounded-degree feature interactions efficiently. DeepFM~\cite{guo2017deepfm} combines the power of Factorization Machines (FM)~\cite{rendle2010factorization} for recommendation and Multi-layer Perceptron (MLP) layers for feature learning. NFM~\cite{he2017neural} introduces the Bi-Interaction pooling operation in neural network modeling. AutoInt~\cite{song2019autoint} applies multi-head self-attention~\cite{Vaswani2017Attention} to automatically learn high-order feature interactions and efficiently handle large-scale high-dimensional sparse data. FiBiNet~\cite{huang2019fibinet} applies the SENet ~\cite{hu2018squeeze} mechanism to dynamically learn the weights of features. AFN~\cite{cheng2020adaptive} proposes the logarithmic transformation layer to learn the power of each feature in a feature combination. DCNv2~\cite{Wang2021DCNv2} leverages low-rank techniques to approximate feature crosses in a subspace for better performance. GDCN~\cite{Wang2023Towards} proposes a gated deep cross network and a field-level dimension optimization approach. 

In this paper, we will show that our compression approach can handle \emph{multiple} deep learning recommendation models and generally achieve better performances than the original models. 

\subsection{Low-Rank Compression for CTR Prediction Models}

CTR prediction models tend to have a large number of parameters and are computationally intensive. To reduce parameter sizes and speed up model inference, a natural idea is to factorize one weight into two or more smaller matrices. There are two pipelines for low-rank decomposing the CTR prediction models, i.e., directly performing matrix decomposition, or first reshaping the weight matrix into a tensor, and then applying tensor decomposition on it. For the matrix decomposition methods, a common technique for low-rank factorization is SVD~\cite{golub2013matrix,anil2022factory}, which can be used in all linear layers.  The dimension of Mixed Dimension (MD) Embeddings~\cite{ginart2021mixed} varies with its query frequency. For the tensor decomposition method, TT-Rec~\cite{yin2021tt} applies Tensor-Train Decomposition (TTD)~\cite{hrinchuk2020tensorized} to compress the embedding layers. Xia et al.~\cite{xia2022device} introduce semi-tensor product based tensor-train decomposition (STTD) for higher compression rates of the embedding table. In previous low-rank approaches such as TT-Rec, a high parameter compression ratio can be reached.

Nevertheless, the flexibility of previous tensor decomposition methods is limited, because it can only be applied to the embeddings. However, the most computation-intensive module in DLRMs is the MLP layer. Besides, it introduces abundant extra computational overhead and takes a too long time for inference. By contrast, our unified framework is the \emph{first} attempt to simultaneously decompose both embedding tables and MLP weights and achieve higher speeds by low-rank compression in CTR prediction models. 


\subsection{Other Compression Methods for CTR Prediction Models}

Parameter pruning is another useful technique for striking a balance between model accuracy and inference speed by cutting out redundant parameters. Plug-in Embedding Pruning (PEP)~\cite{liulearnable} obtains a mixed-dimension embedding scheme by adaptively learning pruning threshold from data. UMEC~\cite{shen2021umec} formulates the joint input feature selection and model compression task as a constrained optimization problem. Then it solves the DLRMs compression task by the alternating direction method of the multipliers algorithm.  SSEDS~\cite{qu2022single} proposes a single-shot embedding pruning method. It first pre-trains a traditional CTR prediction model with unified embedding dimensions. Then it utilizes the proposed criterion which could measure the importance of embedding dimensions only in one forward-backward pass. Besides network pruning, quantization and hashing methods are also commonly used compression recommendation system methods. Product quantization~\cite{Herve2011Product} decomposes the space into a Cartesian product of low-dimensional subspaces and quantizes each subspace separately. Random Offset Block
Embedding (ROBE)~\cite{desai2022random} uses hash functions on embedding tables to locate it in a small circular array of memory. Binary Hash (BH)~\cite{Yan2021Binary} uses a binary code based hash embedding method to reduce the size of the embedding table in arbitrary scale.

Those methods have achieved a high compression ratio in recommendation systems and our approach may be potentially combined with them. Note that though some modern CTR prediction model compression algorithms can achieve higher compression ratios, they often bring a degree of performance degradation, which leads to a large decline in industry revenue. Therefore, if there is no suitable compression algorithm that can improve AUC, algorithm engineers often do not compress the recommended models.


\section{Methods}\label{sec-framework}

In this section, we first describe our framework, starting by introducing traditional matrix and tensor decomposition methods. Then we design two different low-rank approximation algorithms for introducing AFM into the embedding tables and the MLP modules of CTR prediction models, respectively. Throughout our compression process, both algorithms can generally improve the AUC with both fewer parameters and higher speeds. 

\subsection{Classical Low-Rank Compression Methods}

Traditional low-rank decomposition methods, including matrix and tensor decomposition, are widely used in model compression tasks. The most classical method of matrix decomposition is SVD. Specifically, let us consider a matrix $M \in\mathbb{R}^{d_1 \times d_2}$, a typical way to compress this matrix is to perform SVD on $M$, i.e., $M = U S V^\top$, where $U \in\mathbb{R}^{d_1 \times d_1}$ and $V \in\mathbb{R}^{d_2 \times d_2}$  are orthonormal matrices. $S \in\mathbb{R}^{d_1 \times d_2}$ is a diagonal rectangular matrix containing singular values in the decreasing order. If we only use the largest $k$ terms of the singular values, the resulting matrix is an optimal approximation of $M$ with a lower rank $k < \min{(d_1, d_2)}$: $M \approx M_1 M_2$, where $M_1 \in \mathbb{R}^{d_1 \times k}$ and $M_2 \in\mathbb{R}^{k \times d_2}$ are the rank-$k$ approximation matrices by taking $M_1 = U S_{k}^{\frac{1}{2}}$ and $M_2 = S_{k}^{\frac{1}{2}} V^\top$, and $S_{k}^{\frac{1}{2}}$ is a diagonal matrix formed by the square-roots of the corresponding top $k$ singular values in $S$. After this low-rank approximation, the number of parameters in this matrix decreases from $O(d_1 d_2)$ to $O((d_1+d_2)k)$. 

Similar to matrix decomposition, Tensor-Train Decomposition (TTD) is a simple and robust approach to decompose tensor representation of multidimensional data into a product of smaller tensors. Assume a tensor $T \in\mathbb{R}^{d_1 \times d_2 \times \cdots \times d_n }$ is a $n$-dimensional tensor, then we can apply TTD on $T$, i.e., $T \approx T_1 T_2 \cdots T_n$, where $T_i \in\mathbb{R}^{r_{i-1} \times d_i \times r_{i} }$ and $r_0 = r_n = 1 $ to keep the product of the sequence of tensors a scalar. The sequence ${\{r_i\}}^n_{i=0}$ is referred to as TT-ranks, and each 3-dimension tensor $T_i$ is called a TT-core.

The TTD method can also be generalized to compress a matrix $M \in\mathbb{R}^{m \times n}$. We assume that $m$ and $n$ can be factorized into sequences of integers, i.e., $m = \prod_{i=1}^{k}m_i$ and $n = \prod_{i=1}^{k}n_i$. Correspondingly, we reshape the matrix $M$ as a $2n$-dimensional tensor $M' \in R^{(m_1\times n_1) \times (m_2\times n_2) \cdots (m_k\times n_k)}$. Then $M' \approx M'_1 M'_2 \cdots M'_k$, where $M'_i \in\mathbb{R}^{r_{i-1} \times m_i \times n_i  \times r_{i} }$ and $r_0 = r_n = 1 $. Let $\bar{r}$, $\bar{m}$, and $\bar{n}$ be the maximal values of sequences $\{r_i\}$, $\{m_i\}$ and $\{n_i\}$ for $i$ in $ \{1, \cdots ,d\}$, then TTD reduces the space for storing the matrix from $O(mn)$ to $O(k\bar{r}^2\bar{m}\bar{n})$. Please note that in many practical scenarios, it is difficult to find a suitable sequence $\{m_i\}$ and $\{n_i\}$ to accurately decompose $m$ and $n$, so researchers often add some extra rows and columns to M and then perform TTD compression. 

To the best of our knowledge, low-rank decomposition has \emph{not} been fully studied in the area of compressing CTR prediction models. Traditional matrix decomposition methods, such as SVD, have rarely been explored in the field of compressing recommendation systems. Besides, traditional tensor decomposition methods (e.g., TTD) are difficult to apply in MLP layers because they will damage the original structure of the MLP weight, which causes trouble for inference and leads to lower inference speed. In addition, the MLP layer tends to have fewer parameters, and using tensor decomposition methods is not helpful in this aspect either. Besides, embedding table lookup is originally a low computational cost operation. However, to obtain the embeddings, TTD needs to collect all tensors and recalculate embedding tables again. Therefore, although TTD can heavily reduce the model's parameter sizes (e.g., 100$\times$)~\cite{yin2021tt}, it will greatly increase the inference overhead, which limits its practicality. These pressing difficulties prompt us to come up with novel low-rank compression solutions for CTR prediction tasks.

\subsection{Atomic Feature Mimicking for MLP Layers}

Let us consider a fully-connected layer $y = W x + b$ in CTR prediction  models, whose $x \in\mathbb{R}^{n \times c}$, $y \in\mathbb{R}^{m \times c}$ and $W \in\mathbb{R}^{m \times n}$, $b \in\mathbb{R}^{m}$. The optimization target of previous low-rank decomposition algorithms is often aimed at a single matrix or tensor, i.e., $\min \| W - W_r\|^2$, where $W_r$ is the low-rank approximation of $W$. However, compared with only decomposing $W$, we also need to consider the distribution of the input $x$, that is, $\min \|W x - W_r x\|^2$ is a better choice~\cite{yu2020compressing,Hassibi1993Optimal}.

Therefore, following the notation described in~\cite{wu2020essentials,yu2020compressing}, now we introduce Atomic Feature Mimicking (AFM~\cite{yu2020compressing}), which seeks to factorize the output features as opposed to decomposing the model weights. Let us treat the output feature in $\mathbb{R}^{m \times c}$ as $c$ instantiations of the random feature vector $y$ (each in $\mathbb{R}^m$), and compute the covariance matrix:
\begin{equation}
    \operatorname{Cov}(y)=\mathbb{E}\left[y y^{\top}\right]-\mathbb{E}[y] \mathbb{E}[y]^{\top} \,,
    \label{eq:cal_cov_matrix}
\end{equation}
where $\mathbb{E}[\cdot]$ is the expectation operator. Since $\operatorname{Cov}(y)$ is positive semi-definite, its eigendecomposition (i.e., the principal component analysis or PCA) is $\operatorname{Cov}(y)= U S U^\top$. We only keep the top $k$ eigenvalues and extract the first $k$ columns of $U\in \mathbb{R}^{m \times m}$ into $U_{k} \in \mathbb{R}^{m \times k}$ and $U_k U_k^\top \approx I$. The classic PCA knowledge tells us the low-rank representation of $y$ is $U_k^\top ( y - \mathbb{E}[y])$, and $y$ can be approximated as $\mathbb{E}[y] + U_k U_k^\top ( y - \mathbb{E}[y])$. Hence, 
\begin{align}
     y - \mathbb{E}[y] & \approx U_k U_k^\top( y - \mathbb{E}[y])\,,\text{ or,}  \\
     y & \approx U_k U_k^\top y  + \mathbb{E}[y] - U_k U_k^\top \mathbb{E}[y]\,.
\end{align}
This approximation is proved optimal~\cite{wu2020essentials}. Then, one linear layer can be transformed into two: 
\begin{align}
     y & \approx U_kU_k^\top ( W x + b)  + \mathbb{E}[y] - U_k U_k^\top \mathbb{E}[y]\,,\\
       & = U_k ( U_k^\top W x )+ \mathbb{E}[y] + U_k U_k^\top(b - \mathbb{E}[y])\,,
\end{align}
where the first FC layer has weights $U_k^\top W \in \mathbb{R}^{k \times n}$, and the second one has weights $U_k \in \mathbb{R}^{m \times k}$ and bias $\mathbb{E}[y] + U_k U_k^\top(b - \mathbb{E}[y])   \in \mathbb{R}^{m}$. Note that because collecting output $y$ during inference would be a memory-greedy process, in practice we adaptively update $\mathbb{E}[yy^\top]$ and $\mathbb{E}[y]$ in a streaming fashion instead of storing all output features.  We will perform inference on the entire training set, and then collect the $\mathbb{E}[yy^\top]$ and $\mathbb{E}[y]$ of each FC layer in the MLP of the recommendation system to compress them.

Furthermore, surprisingly, our subsequent experiments will show that for the MLP module in CTR prediction models, after initializing the compressed model parameters by AFM, adding an extra activation function between $U_k^TW$ and $U_k$ only slightly degrades AUC. Moreover, this strategy can obtain higher AUC scores after fine-tuning. Therefore, we improve AFM by adding ReLU~\cite{nair2010rectified} activation functions into the model structure after compression and then fine-tune the whole compressed model for 1 epoch.

For CTR prediction models, compared with traditional matrix decomposition methods (such as SVD), because AFM imitates the output $y$ rather than the weight $W$, the AUC drop it brings will be much smaller. Therefore, AFM is a better matrix-decomposition initialization method. Compared with traditional tensor decomposition methods like TTD, AFM obtains higher speed acceleration because it requires fewer multiplication calculations.

\subsection{Compressing the Embedding Tables}

The embedding tables map every sparse categorical feature to real-valued dense vectors. In this subsection, we will show how AFM can be made to compress the embedding tables.

Let us denote an input vector as $v \in \mathbb{R}^d$, in which $v_i$ is its $i$-th feature field. $v_i$ can be either a continuous or nominal variable. For notational simplicity, we assume $v_i$ has an embedding $\mathbf{e}_i$. When $v_i$ is continuous, $\mathbf{e}_i = v_i \in \mathbb{R}$ and it does not require embedding tables, so we do not consider continuous variables. When it is nominal, $v_i$ is in fact an index value and there is an embedding table $D_i \in \mathbb{R}^{t \times s_i}$ associated with $v_i$. Here both $t$  and $s_i$ are constant values, i.e., $t$ is the embedding dimensionality shared by all fields, and $s_i$ is the number of possible items in the dictionary such that $v_i \in \{1, 2, \cdots, s_i\}$. Normally $D_i$ is a dense matrix, and $\mathbf{e}_i =  D_i(:, v_i) \in \mathbb{R}^{t \times 1}$. Now the output $y$ of the embedding layer is 
\begin{equation}
    \begin{aligned}
    y  =\left[\mathbf{e}_1, \mathbf{e}_2, \ldots, \mathbf{e}_d \right]\,. \\
    \end{aligned}
\end{equation}

The nominal feature embedding layer $D = \{D_1, D_2, \cdots, D_d\}$ occupies most of the parameters in the models and now our goal is to compress the embedding dimensionality $t$. Here, we compute the covariance matrix independently for each single categorical field. Let us consider the embedding table $D_i$. In particular, the embedding table lookup operation also can be regarded as a fully-connected layer, i.e., $\mathbf{e}_i =  D_i(:, v_i) = D_i x_{v_i}$. Here $x_{v_i} \in \mathbb{R}^{s_i \times 1}$ is a one-hot vector, i.e., the $v_i$-th element of  $x_{v_i}$ is 1, and the others are 0. Therefore, we can perform atomic feature mimicking on the embedding table $D_i$. First, we collect embedding output $\mathbf{e}_i$ in the whole training dataset. Then we calculate $\mathbb{E}[\mathbf{e}_i]$ and $\mathbb{E}[\mathbf{e}_i \mathbf{e}_i^T]$. After that, we compute the covariance matrix of $\mathbf{e}_i$:
\begin{equation}
    \operatorname{Cov}(\mathbf{e}_i)=\mathbb{E}\left[\mathbf{e}_i \mathbf{e}_i^{\top}\right]-\mathbb{E}[\mathbf{e}_i] \mathbb{E}[\mathbf{e}_i]^{\top} =  U^i S^i (U^i)^\top\,.
\end{equation}

Similarly, we extract the first $k$ columns of  $U^i\in \mathbb{R}^{t \times t}$ into $U^i_{k} \in \mathbb{R}^{t \times k}$. Therefore, we can get 
\begin{equation}
\mathbf{e}_i - \mathbb{E}[\mathbf{e}_i] \approx U^i_{k} (U^i_{k})^\top( \mathbf{e}_i - \mathbb{E}[\mathbf{e}_i])\,.
\end{equation}

Hence, $\mathbf{e}_i = D_i x_{v_i}$ can be approximated as
\begin{align}
    \mathbf{e}_i & \approx U^i_{k} (U^i_{k})^\top( D_i x_{v_i} - \mathbb{E}[\mathbf{e}_i]) + \mathbb{E}[\mathbf{e}_i]\,,\\
      & = U^i_{k} ( (U^i_{k})^\top D_i x_{v_i} )+ (I - U^i_{k}(U^i_{k})^\top) \mathbb{E}[\mathbf{e}_i]\,.
\end{align}

Therefore, the $i$-th embedding table $D_i \in \mathbb{R}^{ t \times s_i}$ can be replaced by $ (U^i_{k})^\top D_i \in \mathbb{R}^{ k \times s_i}$. For other embedding tables, we can also perform similar operations. Then the embedding dimension of the entire recommendation model will be reduced from $t$ to $k$. Now any input vector $v$ has its compressed embedding $\mathbf{e}' = [\mathbf{e}'_1 , \mathbf{e}'_2, \cdots, \mathbf{e}'_d]$, where $\mathbf{e}' \in \mathbb{R}^{\sum_{i=1}^d k}$.

It is worth noting that after compressing embeddings, we add an extra fully-connected layer after each embedding table. For the $i$-th embedding table, the weight is $W_i = U^i_{k} \in \mathbb{R}^{ t \times k}$ and bias is $ b_i = (I - U^i_{k}(U^i_{k})^\top) \mathbb{E}[\mathbf{e}_i]\in \mathbb{R}^{ t}$.  Therefore, the final output $y^e$ of the embedding layer after AFM is  
\begin{equation}
    y^e  =\left[W_1\mathbf{e}'_1 + b_1, W_2\mathbf{e}'_2 + b_2, \ldots, W_d\mathbf{e}'_d + b_d\right] \,,
\end{equation}
where $y_e \in \mathbb{R}^{\sum_{i=1}^d t}$. Hence, we reduce the space for storing the embedding tables from $O(t \sum_{i=1}^{d} s_i)$ to $O(k (\sum_{i=1}^{d} s_i + dt))$.  It is worth noting that $W_i$ is often very lightweight, which means the cost of this calculation can be ignored.

In particular, the layers after the embedding tables are usually an MLP layer and a feature interaction module. Then, to obtain a higher compression ratio, we can further merge the first linear layer in the MLP and the linear layer brought by compressing embedding tables. In particular, we express the computation of the first linear layer in the MLP as $y' = W'x+b'$. Suppose the output of this layer has $o$ dimensions, then $W' \in \mathbb{R}^{o \times \sum_{i=1}^{d}t}$ and $b' \in \mathbb{R}^{o}$. First, let us rewrite $W'$ in the block matrix notation, i.e., $W' = [W'_1,W'_2, \cdots ,W'_d]$, in which $W'_i \in \mathbb{R}^{o \times t}$. Besides, we can also rewrite $y^e$ as $[y^e_1, y^e_2, \cdots, y^e_d]$, where $y^e_i = W_i\mathbf{e}'_i + b_i$. Then, it is obvious that the input $x$ of the first MLP layer $y' = W'x+b'$ is $y^e$, i.e., 
\begin{align}
        y' &= W'y^e+b' \notag\\
    &=  [W'_1 y^e_1,W'_2 y^e_2, \cdots ,W'_d y^e_d]+b' \notag\\
    &=  [W'_1 (W_1\mathbf{e}'_1 + b_1),W'_2 (W_2\mathbf{e}'_2 + b_2), \cdots ,W'_d (W_d\mathbf{e}'_d + b_d)]+b' \notag\\
    &=   [W'_1 W_1\mathbf{e}'_1,W'_2 W_2\mathbf{e}'_2, \cdots ,W'_d W_d\mathbf{e}'_d] \notag\\
    &\phantom{=} + [W'_1 b_1, W'_2 b_2, \cdots, W'_d b_d] + b'.
\end{align}

Then we can replace $W'$ with $[W'_1 W_1 ,W'_2 W_2 , \cdots ,W'_d W_d ]$, and $b'$ with $[W'_1 b_1, W'_2 b_2, \cdots, W'_d b_d] + b'$. After this replacement, the size of the first dense layer (mostly in $W'$) in the MLP layers will be reduced, too. Although this reduction is negligible when compared to the reduction in the embedding tables, it will compress the input dimension size of $W'$ from $ \sum_{i=1}^{d}t$ to $ \sum_{i=1}^{d}k$ and will further improve the inference speed. Note that at this time we still need to retain $[W_1, W_2, \cdots, W_d ]$ to calculate the embedding output and use this as the input of the feature interaction module. Moreover, for some specific models (such as FiBiNet~\cite{huang2019fibinet}), which will add an additional interaction layer between the embedding tables and the MLP layers, this fusion method is not applicable.

After compressing embedding tables, we also fine-tune the whole compressed model for 1 epoch. Later we will show that after fine-tuning, the compressed model's AUC is also significantly better than the original model. It is obvious that our two compression strategies can be jointly applied and achieve higher inference speed with lower parameter sizes.

In total, our entire compression process is as follows. First, we use AFM to compress the MLP module of the CTR prediction models and fine-tune the models by one epoch. Then we continue to reduce the embedding dimensionality based on the previous compressed model. According to the structure of each CTR prediction model, we judge whether to merge the projection weight of embeddings into the first linear layer of MLP or not. Note that throughout the compression process, we apply the whole training dataset to calculate the weights of AFM and fine-tune the models. 

It can be seen that our plug-and-play framework has very little intrusiveness to the entire recommendation system. After compressing the MLP layers, without making any changes to the codebase, algorithm engineers only need to add some extra fully-connected layers to the model weights. After further compressing the embedding, only a few lightweight dense layers need to be added after the embedding tables. These features make our framework very easy to implement and user-friendly. Later we will demonstrate that our framework can attain even higher AUC with significantly fewer parameters and faster inference speed on these tasks.  

\section{Experiments}

In this section, we start by describing the datasets, evaluation metrics, and baseline models used for our experiments. Then, we perform our methods on those pre-trained models and compare our algorithms to state-of-the-art previous low-rank decomposition works. We also list several ablation studies and end this section with online experiments. All the experiments are conducted with PyTorch. More training details and further experimental results are shown in the appendix.

\subsection{Datasets, Metrics and Models}

\textbf{Datasets.} We evaluate the proposed method on two public datasets for CTR prediction tasks, i.e., Criteo and Avazu. Criteo contains click logs with 45.8 million data instances, and Avazu consists of several days of 40.4 million ad click-through data which is ordered chronologically. Besides, we also evaluate our algorithms on a private industry dataset, i.e., the XYZ AppGallery dataset.\footnote{Please note that the company name is anonymized as XYZ.} Its training dataset consists of user click information on the XYZ AppGallery within 12 days, and its testing dataset contains user click logs on the 13th day. We follow the hyperparameter settings in FiBiNet~\cite{huang2019fibinet}, i.e., we split the Criteo dataset randomly into two parts: 90\% is for training, while the rest is for testing.  We also split Avazu randomly into two parts: 80\% is for training, while the rest is for testing. We shown the statistics of the Criteo, Avazu and XYZ AppGallery datasets in Table~\ref{tab:data_info}.


\begin{table}
    \centering
    \small
    \caption{Statistics of the three recommendation datasets used in our experiments.}
    \label{tab:data_info}
    \begin{tabular}{l|cccc}
    \bottomrule[1pt]
    Name    & \# Train &  \# Test & \# Categorical  & \# Continuous  \\ \Xhline{2\arrayrulewidth}
    Criteo   & 41256556 & 4584062 & 26 & 13\\  
    Avazu    & 32343174 & 8085794 & 22 & 0 \\ 
    XYZ  & 10842707 & 20538 & 80 & 0\\ 
    \toprule[1pt] 
    \end{tabular}
\end{table}

\textbf{Evaluation Metrics.} We adopt AUC (Area Under the ROC Curve) and Logloss to measure the performance of models. We also report the parameter sizes of the recommendation models and average inference throughput in the test dataset. Because the inference bottleneck of the recommendation systems is memory access, to show the benefits brought by our compressing algorithms clearly, we calculate the models' throughput on an Intel Xeon Gold 5220R CPU with a fixed 10000 mini-batch size. Note that an improvement of 1\textperthousand~in AUC is usually regarded as significant for CTR prediction tasks, because it will bring a large increase in a company's revenue if the company has a substantial user base.

\textbf{Baseline models.} To validate the effectiveness of our compression method, we deploy our framework to seven representative CTR prediction models: DCN~\cite{wang2017deep}, DeepFM~\cite{guo2017deepfm}, NFM~\cite{he2017neural}, AutoInt~\cite{song2019autoint}, FiBiNet~\cite{huang2019fibinet}, DCNv2~\cite{Wang2021DCNv2} and GDCN~\cite{Wang2023Towards}. To show the influences of AFM, we compare it with recent advances in low-rank approximation: SVD~\cite{golub2013matrix} and TTD~\cite{yin2021tt}.  

\subsection{Main Experiments}\label{sec:compressing_the_dlrms}

We summarize the performances on Criteo and Avazu test sets in Table~\ref{tab:compress_dlrm_criteo_avazu}, and show the results on XYZ AppGallery test set in Table~\ref{tab:compress_dlrm_XYZ}. 

\begin{table*}[t]
    \centering
    \small
    \caption{Results on the Criteo (columns 2--5) and Avazu (columns 6--9) test sets.}
    \label{tab:compress_dlrm_criteo_avazu}
    \begin{tabular}{l|cccc||cccc}
    \bottomrule[1pt]
    Model    & AUC & Logloss   & Param. (M)  &   Throughput    & AUC & Logloss   & Param. (M) &   Throughput  \\ \Xhline{2\arrayrulewidth}
    DCN   & 0.7932 & 0.4570 & 574.46  & 41989.02 & 0.7890 & 0.3745 &  492.14  & 4412.13\\  \hline
    \quad+AFM MLP  & 0.8013  & 0.4496  & 574.24 & 50468.93 (+20.2\%) & 0.7933 & 0.3719 &  486.70  & 7013.75 (+59.0\%)\\
    \quad+AFM EMB & \textbf{0.8023} & \textbf{0.4489} &  \textbf{101.42}  & \textbf{58661.30} (+39.7\%) & \textbf{0.7941} & \textbf{0.3715} &  \textbf{87.98}  & \textbf{10203.356} (+131.3\%)\\ \Xhline{2\arrayrulewidth}
    DeepFM   & 0.7964  & 0.4541 & 574.46   & 46094.45 & 0.7917  & 0.3728 & 492.13   & 4312.46 \\  \hline
    \quad+AFM MLP  & 0.8016 & 0.4498 & 574.24 & 59877.58 (+29.9\%) &\textbf{0.7964}  & \textbf{0.3699}    & 486.69 & 7417.67 (+72.0\%)\\
    \quad+AFM EMB & \textbf{0.8021} & \textbf{0.4495} & \textbf{101.42} & \textbf{73365.64} (+59.2\%)&  0.7948&  0.3716  & \textbf{87.98} & \textbf{11474.46} (+166.1\%)\\ \Xhline{2\arrayrulewidth}
    NFM    & 0.7929     & 0.4571 &  574.30  & 58398.40 & 0.7864  & 0.3764 &  490.03  &  5388.11 \\ \hline
    \quad+AFM MLP  & 0.8016 & 0.4492 & 574.08 & 77259.30 (+32.3\%)& \textbf{0.7917} & \textbf{0.3729} & 484.59    & 11792.61 (+118.9\%)\\
    \quad+AFM EMB & \textbf{0.8025} & \textbf{0.4484} & \textbf{101.26} & \textbf{80218.91} (+37.4\%)& 0.7911 &  0.3738  & \textbf{85.87}  & \textbf{11835.38} (+119.7\%)\\ \Xhline{2\arrayrulewidth}
    AutoInt     & 0.7939 & 0.4563 & 574.46 & 11637.07 & 0.7904 &  0.3735     & 492.16 &  3496.45 \\  \hline
    \quad+AFM MLP  & 0.8016 & 0.4496 & 574.24 & 12101.31 (+3.99\%)& \textbf{0.7957}  & \textbf{0.3705}  & 486.72 & 4339.63 (+24.1\%)\\
    \quad+AFM EMB & \textbf{0.8019} & \textbf{0.4494}  & \textbf{101.42} & \textbf{12508.22} (+7.49\%)& 0.7952 & 0.3708 & \textbf{88.00} & \textbf{5901.96} (+68.8\%)\\ \Xhline{2\arrayrulewidth}
    FiBiNet     & 0.8002 & 0.4509 & 578.62 & 3989.23 & 0.7965 & 0.3699   & 537.29 &  612.23 \\  \hline
    \quad+AFM MLP  & \textbf{0.8055}  & \textbf{0.4458} & 578.40 & 
    4070.20 (+2.03\%) & \textbf{0.8011}  & \textbf{0.3668}   & 531.85 & 673.45 (+10.0\%)\\
    \quad+AFM EMB & 0.8051 & 0.4462 & \textbf{105.58} & \textbf{4552.54} (+14.12\%)& 0.7968 & 0.3699  & \textbf{133.13} & \textbf{713.36} (+16.5\%)\\ \Xhline{2\arrayrulewidth}
    DCNv2     &0.7947 & 0.4562	&574.83 &	32928.05 & 0.7913 & 	0.3731 & 	494.55  & 	3691.68 \\  \hline
    \quad+AFM MLP  & 0.8026 & 0.4486  &  574.61  & 	40466.83 (+22.89\%)	 & \textbf{0.7959} &	\textbf{0.3704}	 & 489.11 &	6057.72 (+64.09\%)\\
    \quad+AFM EMB &\textbf{ 0.8029} & 	\textbf{0.4485} & 	\textbf{101.79} & 	45162.29 (+37.15\%) & 	0.7950 & 	0.3709 & 	\textbf{90.39} & 	\textbf{7968.40} (+115.85\%) \\ \Xhline{2\arrayrulewidth}
    GDCN     &0.7949	 & 0.4559 & 	575.19 & 	26267.60 & 	0.7913	 & 0.3729 & 	496.97	 & 3099.29
 \\  \hline
    \quad+AFM MLP  & 0.8015 & 	0.4496 & 	574.97	 & 28574.70 (+8.78\%) & 	\textbf{0.7962} & \textbf{	0.3701}	 & 491.53	 & 4564.05 (+47.26\%)\\
    \quad+AFM EMB &\textbf{0.8022} & 	\textbf{0.4491}	 & \textbf{102.15}	 &\textbf{ 29113.02 }(+10.83\%)	 & 0.7953	 & 0.3706	 & \textbf{92.81} & \textbf{5582.37} (+80.12\%)\\ \Xhline{2\arrayrulewidth}

    \toprule[1pt] 
    \end{tabular}
\end{table*}

\begin{table}
    \centering
    \small
    \caption{Results on the XYZ AppGallery test set.}
    \label{tab:compress_dlrm_XYZ}
    \begin{tabular}{l|cccc}
    \bottomrule[1pt]
    Model    & AUC & Logloss   & Param. (M)  &   Throughput     \\ \Xhline{2\arrayrulewidth}
    DCN   & 0.8271 & 0.1373 & 633.61 & 18810.37 \\  \hline
    \quad+AFM MLP     & 0.8299 & 0.1378 & 633.39 & \textbf{20067.73} (+6.68\%)  \\
    \quad+AFM EMB    & \textbf{0.8310} & \textbf{0.1363} & \textbf{186.35} & 19259.32 (+2.39\%) \\ \Xhline{2\arrayrulewidth}
    DeepFM & 0.8290 & 0.1364 & 633.61 & 31746.22 \\  \hline
    \quad+AFM MLP    & 0.8306 & 0.1381 & 633.39 & 34462.58 (+8.56\%) \\
    \quad+AFM EMB    & \textbf{0.8362} & \textbf{0.1356} & \textbf{186.35} & \textbf{39714.90}  (+25.1\%) \\ \Xhline{2\arrayrulewidth}
    NFM       & 0.8330 & 0.1370 & 633.10 & 47358.46 \\ \hline
    \quad+AFM MLP    & 0.8300 & 0.1369 & 632.88  & \textbf{55643.18} (+17.5\%)  \\
    \quad+AFM EMB    & \textbf{0.8336} &	\textbf{0.1359} & \textbf{186.23} & 52233.17 (+10.3\%) \\ \Xhline{2\arrayrulewidth}
    AutoInt       & 0.8266 & 0.1371 & 633.61 & 2644.54  \\  \hline
    \quad+AFM MLP    & \textbf{0.8348} & \textbf{0.1358} & 633.39 & \textbf{2950.49} (+11.6\%) \\
    \quad+AFM EMB    & 0.8312 &	0.1365 & \textbf{186.35} & 2805.43 (+6.08\%) \\ \Xhline{2\arrayrulewidth}
    FiBiNet        & 0.8341 & \textbf{0.1350} & 675.16 & 446.49  \\  \hline
    \quad+AFM MLP    & 0.8326 & 0.1363 & 674.95 & 464.86 (+4.11\%) \\
    \quad+AFM EMB    & \textbf{0.8372} &	0.1356 & \textbf{228.29} & \textbf{502.01} (+12.4\%) \\ \Xhline{2\arrayrulewidth}
    DCNv2        & 0.8339 &	0.1350 &	636.89	 &8382.67 \\  \hline
    \quad+AFM MLP    & 0.8343 &	0.1349	 &636.67 &	8882.50 (+5.96\%) \\
    \quad+AFM EMB    & \textbf{0.8347} &	\textbf{0.1344}	 &\textbf{189.63} &	\textbf{8965.79} (+6.96\%) \\ \Xhline{2\arrayrulewidth}
    GDCN        & 0.8315& 	0.1358& 	640.16& 	4970.88\\  \hline
    \quad+AFM MLP    & \textbf{0.8397}	& \textbf{0.1347}& 	639.94& 	5195.75 (+4.52\%) \\
    \quad+AFM EMB    & 0.8348&	0.1348	&\textbf{193.12}&	\textbf{5265.81 (+5.93\%)} \\ \Xhline{2\arrayrulewidth}

    \toprule[1pt] 
    \end{tabular}
\end{table}

\textbf{Implementation details.} We first train the baseline models with uniform embedded dimensions. For the Criteo and Avazu datasets, we follow the model hyperparameters settings in FiBiNet, i.e., we set 400 dimensions per layer for the Criteo dataset and 2000 neurons per layer for the Avazu dataset. The embedding dimension is 16 for the Criteo dataset and 50 for the Avazu dataset. For the XYZ AppGallery dataset, we also set the MLP hidden sizes to 400 and the embedding dimension to 16. For all models, the number of hidden layers is set to 3 and we apply the ReLU activation function. In particular, we also add a linear weight for each categorical field, whose input and output dimensions are the number of possible items and 1, respectively. 

Then we compress these base models by our framework. We first compress the MLP layer and set the compression dimensions to 64 and 320 for the Criteo and Avazu datasets, respectively. For example, on the Criteo dataset, we low-rank decompose the original 400$\times$400 linear layer into a 400$\times$64 and a 64$\times$400 linear layer. For the XYZ AppGallery dataset, the compression dimension is 64, too. Then we apply the entire training set to fine-tune the compressed model for 1 epoch. We only compress the second and third linear layers and do not deal with the first layer. This is because the input dimension of the first dense layer is related to the embedded layer, and in some cases, the input dimension is shallow. For simplicity, we do not consider the first dense linear for all models.

Then we continue to reduce the embedding dimensions of those MLP-compressed models. For the Criteo and Avazu datasets, we uniformly compress the embedding dimension from 16 to 2, and 50 to 8, respectively. For the XYZ AppGallery dataset, we reduce the dimension from 16 to 4. After compressing the embedding dimension, we still fine-tune the sub-model for 1 epoch. When compressing NFM and FiBiNet, due to their unique structure, we do not merge the compression weights into the first dense layer in the MLP module.
\begin{table*}[t]
    \centering
    \small
    \caption{Results of AFM, SVD, and TTD on the Criteo (columns 3--6) and Avazu (columns 7--10) test sets with DeepFM.}
    \label{tab:ablation_svd_afm_ttd}
    \begin{tabular}{ll||cc|c|c||cc|c|c}
    \bottomrule[1pt]
    \multicolumn{2}{l||}{DeepFM}                       &AUC  & Logloss & Param. (M) & Throughput &AUC  & Logloss & Param. (M) & Throughput\\ \hline
    \multicolumn{2}{l||}{Baseline} & 0.7964  & 0.4541 & 574.46   & 46094.45 & 0.7917  & 0.3728 & 492.13   & 4312.46 \\  \Xhline{2\arrayrulewidth}
    \multicolumn{1}{l|}{\multirow{2}{*}{\quad+AFM MLP}} & Pre-train  & 0.7964 & 0.4541 & \multirow{2}{*}{574.24} & \multirow{2}{*}{59877.58 (+29.9\%)}  & 0.7913 & 0.3750 & \multirow{2}{*}{486.69}  & \multirow{2}{*}{7417.67 (+72.0\%)}\\ \cline{2-4} \cline{7-8} 
    \multicolumn{1}{l|}{}                         & Fine-tune & 0.8016 & 0.4498  &  & & \textbf{0.7964} & \textbf{0.3699} &  &\\ \hline
    \multicolumn{1}{l|}{\multirow{2}{*}{\quad+AFM EMB}} & Pre-train  & 0.7900 & 0.4601 &  \multirow{2}{*}{101.42}  & \multirow{2}{*}{\textbf{73365.64} (+59.2\%)}& 0.7942 & 0.3720 &  \multirow{2}{*}{87.98}  & \multirow{2}{*}{\textbf{11474.46} (+166.1\%)}\\ \cline{2-4} \cline{7-8} 
    \multicolumn{1}{l|}{}                         & Fine-tune  &\textbf{0.8021} & \textbf{0.4495}&  & & 0.7948 & 0.3716 &  &\\  \Xhline{2\arrayrulewidth}
    \multicolumn{1}{l|}{\multirow{2}{*}{\quad+SVD MLP}} & Pre-train  & 0.7352 & 0.7296 & \multirow{2}{*}{574.24} & \multirow{2}{*}{59877.58 (+29.9\%)} & 0.7643 & 0.4759 &  \multirow{2}{*}{486.69}  & \multirow{2}{*}{7417.67 (+72.0\%)}\\ \cline{2-4} \cline{7-8} 
    \multicolumn{1}{l|}{}                         & Fine-tune  & 0.8011 & 0.4503 &  & & 0.7960   & 0.3702  &  &\\ \hline
    \multicolumn{1}{l|}{\multirow{2}{*}{\quad+SVD EMB}} & Pre-train  & 0.7833 & 0.4761 &  \multirow{2}{*}{101.42}  & \multirow{2}{*}{\textbf{73365.64} (+59.2\%)} & 0.7930 & 0.3734 & \multirow{2}{*}{87.98}  & \multirow{2}{*}{\textbf{11474.46} (+166.1\%)}\\ \cline{2-4} \cline{7-8} 
    \multicolumn{1}{l|}{}                         & Fine-tune  & 0.8009 & 0.4503 &  & & 0.7933 & 0.3732 &  &\\  \Xhline{2\arrayrulewidth}
    \multicolumn{1}{l|}{\multirow{2}{*}{\quad+TTD EMB}} & Pre-train  & 0.7667 & 0.4801 & \multirow{2}{*}{\textbf{3.76}}  & \multirow{2}{*}{494.00 (-98.93\%)} & 0.7555 & 0.3938 & \multirow{2}{*}{\textbf{12.21}} & \multirow{2}{*}{1318.03 (-69.44\%)}\\ \cline{2-4} \cline{7-8} 
    \multicolumn{1}{l|}{}                         & Fine-tune  & 0.7837 & 0.465 &  & & 0.7644 & 0.3880 &  &\\     \toprule[1pt] 
    \end{tabular}
\end{table*}

\begin{table}[t]
    \centering
    \small
    \caption{Results of AFM, SVD, and TTD on the XYZ AppGallery test set with DeepFM.}
    \label{tab:ablation_svd_afm_ttd_XYZ}
    \begin{tabular}{ll|cc|c|c}
    \bottomrule[1pt]
    \multicolumn{2}{l|}{DeepFM}                       &AUC  & Logloss & Param. & Throughput \\ \hline
    \multicolumn{2}{l|}{Baseline} &  0.8290 & 0.1364 & 633.61 & 31746.22  \\  \Xhline{2\arrayrulewidth}
    \multicolumn{1}{l|}{\multirow{2}{*}{~~+AFM MLP}} & PT  & 0.8290 & 0.1365 & \multirow{2}{*}{633.39} & \multirow{2}{*}{34462.58  (+8.56\%)} \\ \cline{2-4}  
    \multicolumn{1}{l|}{}                         & FT & 0.8306 & 0.1381&  & \\ \hline
    \multicolumn{1}{l|}{\multirow{2}{*}{~~+AFM EMB}} & PT  & 0.8322 & 0.1390 &  \multirow{2}{*}{186.35}   & \multirow{2}{*}{\textbf{39714.90} (+25.1\%)}\\ \cline{2-4}  
    \multicolumn{1}{l|}{}                         & FT  & \textbf{0.8362} & \textbf{0.1356} &  & \\  \Xhline{2\arrayrulewidth}
    \multicolumn{1}{l|}{\multirow{2}{*}{~~+SVD MLP}} & PT  & 0.7869 & 0.2302 & \multirow{2}{*}{633.39} & \multirow{2}{*}{34462.58 (+8.56\%)}  \\ \cline{2-4}  
    \multicolumn{1}{l|}{}                         & FT  & 0.8245 & 0.1380 &  & \\ \hline
    \multicolumn{1}{l|}{\multirow{2}{*}{~~+SVD EMB}} & PT  & 0.8239 & 0.1393 &  \multirow{2}{*}{186.35}  & \multirow{2}{*}{\textbf{39714.90} (+25.1\%)} \\ \cline{2-4}  
    \multicolumn{1}{l|}{}                         & FT  & 0.8339 & 0.1362 &  & \\  \Xhline{2\arrayrulewidth}
    \multicolumn{1}{l|}{\multirow{2}{*}{~~+TTD EMB}} & PT  & 0.8302 & 0.1387 & \multirow{2}{*}{\textbf{12.00}}  & \multirow{2}{*}{2141.07 (-93.26\%)} \\ \cline{2-4}  
    \multicolumn{1}{l|}{}                         & FT  & 0.8340 & 0.1359 &  & \\     \toprule[1pt] 
    \end{tabular}
\end{table}

\textbf{Results.}  We name our improved atomic feature mimicking for embedding tables as AFM EMB, and for MLP layers as AFM MLP. After compressing the MLP layers, except the NFM and FiBiNet on the XYZ AppGallery dataset, all models' AUC  increases significantly, and the LogLoss also largely decreases. After further compressing the embedding tables, the parameter sizes of those models are generally reduced by 3-5$\times$, and the AUC also exceeds the original model by 0.010-0.003, which is a very high improvement in CTR prediction tasks. Because each model contains a linear weight, although the embedding dimension is reduced by 6-8$\times$, the total parameter sizes of those models are reduced by only 3-5$\times$. However, it is still a large compression ratio.  Because the attention module of AutoInt and the SENet module of FiBiNet are very time-consuming, although we have achieved a higher parameter compression rate, the throughput improvement of the two models on the Criteo is about 10\%. Except for these two models, our embedding tables and MLP compression methods can increase the inference speed by 35\%-170\%, which is a very significant improvement. On the XYZ AppGallery dataset, our framework also generally achieves a 10\%-30\% speed increase.

%

\subsection{Ablation Studies}

To explore the influence of different modules of our method, we perform three analyses in this section. We take DeepFM on the Criteo, Avazu, and XYZ AppGallery datasets as examples. For a fair comparison, we adopt the same training strategies as in Section~\ref*{sec:compressing_the_dlrms}. 

%

\subsubsection{Compare AFM with TTD and SVD}
First, we explore the advantages of our method compared with the traditional low-rank decomposition algorithm SVD and TTD. For the convenience of comparison, when applying SVD, we apply the same model structure as AFM but use SVD to initialize the compressed model. When performing TTD, we follow the model hyperparameter settings in the TT-Rec paper~\cite{yin2021tt} and denote the TT-ranks as 16.

Table~\ref{tab:ablation_svd_afm_ttd} shows the performances of those compressed models on the Criteo and Avazu datasets. We name the results of direct compression and further fine-tuning 1 epoch as ``Pre-train'' and ``Fine-tune''. We then display the results of the XYZ AppGallery dataset in Table~\ref{tab:ablation_svd_afm_ttd_XYZ}. We refer to ``Pre-train'' as ``PT''  and ``Fine-tune'' as ``FT''.  As we can conclude from those tables, though AFM indeed performs better than SVD, they both achieve better than the original models. This indicates our framework does not hinge on AFM. In fact, it is our low-rank compression framework method as a whole, rather than the single AFM, that is effective in compressing CTR prediction models. We believe that our framework reveals the great potential of matrix decomposition (not just AFM) in compressed CTR prediction models, which is our greatest contribution. In addition, though TTD can achieve a higher parameter compression rate, it will greatly slow down the inference.



\subsubsection{Add Activation Functions When Compressing MLP Layers}

When compressing the MLP layers, we will add an additional ReLU activation layer between the two decomposed linear layers. Here we explore the effect of this ReLU layer on the results. We follow the previous MLP compression settings. 



Table~\ref{tab:ablation_with_without_relu} shows the results of adding additional activation functions. We report the results of direct compression and further fine-tuning. Surprisingly, after compressing, the ReLU function has only a slight impact on the AUC. After fine-tuning, the AUCs of those models with additional ReLU functions are generally higher than those of models without ReLU. 
Therefore, when low-rank decomposing one linear layer into two linear layers, we will add an additional activation function between the two linear layers.
\begin{table*}[t]
    \centering
    \small
    \caption{Results with or without ReLU functions.}
    \label{tab:ablation_with_without_relu}
    \begin{tabular}{ll|cc||ll|cc||ll|cc}
    \bottomrule[1pt]
    \multicolumn{2}{l|}{Criteo}                       &AUC  & Logloss & \multicolumn{2}{l|}{Avazu}  &AUC  & Logloss & \multicolumn{2}{l|}{XYZ}  &AUC  & Logloss \\ \hline
    \multicolumn{1}{l|}{\multirow{2}{*}{w/ ReLU}} & PT  & 0.7964 & 0.4541  & \multicolumn{1}{l|}{\multirow{2}{*}{w/ ReLU}}  & PT  & 0.7913 & 0.3750 & \multicolumn{1}{l|}{\multirow{2}{*}{w/ ReLU}}  & PT  & 0.8290 & 0.1365 \\ \cline{2-4}\cline{6-8}\cline{10-12}
    \multicolumn{1}{l|}{}                        & FT & \textbf{0.8016} & \textbf{0.4498}  & \multicolumn{1}{l|}{}                        & FT & \textbf{0.7964}  & \textbf{0.3699} & \multicolumn{1}{l|}{}                        & FT & \textbf{0.8306}  & \textbf{0.1381}\\ \Xhline{2\arrayrulewidth}
    \multicolumn{1}{l|}{\multirow{2}{*}{w/o ReLU}} & PT  & 0.7964 & 0.4541  & \multicolumn{1}{l|}{\multirow{2}{*}{w/o ReLU}}  & PT  & 0.7917 & 0.3728  & \multicolumn{1}{l|}{\multirow{2}{*}{w/o ReLU}}  & PT  & 0.8290 & 0.1364 \\ \cline{2-4}\cline{6-8}\cline{10-12}
    \multicolumn{1}{l|}{}                        & FT & 0.8011 & 0.4501  & \multicolumn{1}{l|}{}                        & FT & 0.7961  & 0.3701 & \multicolumn{1}{l|}{}                        & FT & 0.8278  &  0.1375\\ \toprule[1pt] 
    \end{tabular}
\end{table*}

\subsubsection{Dimensionality of Embedding and MLP Layers}

In the previous experiments, we fix the compression dimension of those models. In this subsection, we continue to explore the influence of different compression dimensionalities on model performances. Note that unlike previous experiments, we compress the embedding tables or the MLP layers directly on the baseline model.

We first explore the performances of different embedding dimensions. On the Criteo and XYZ AppGallery datasets, the original embedding dimension is 16, and here we compress the dimensions to $\{2, 4, 8, 12\}$. On the Avazu dataset, the original dimension is 50, and we reduce the dimensions into $\{4, 8, 16, 32\}$. The results are shown in Table~\ref{tab:ablation_different_embedding_dimension}. When directly compressing, the more the embedding dimensions, the higher the AUC of the model. After further fine-tuning, as the dimension increases, the AUC of the model on the Criteo and XYZ AppGallery datasets first increases and then decreases slightly. On the Avazu dataset, the AUC keeps increasing. 


\begin{table*}[t]
    \centering
    \small
    \caption{Results of different embedding dimensionalities.}
    \label{tab:ablation_different_embedding_dimension}
    \begin{tabular}{l|cc|cc|cc|cc}
    \bottomrule[1pt]
    Datasets & AUC  & LogLoss & AUC  & LogLoss & AUC  & LogLoss & AUC  & LogLoss \\ \Xhline{2\arrayrulewidth}
    Criteo & \multicolumn{2}{c|}{Emb = 2}   &  \multicolumn{2}{c|}{Emb = 4}  &  \multicolumn{2}{c|}{Emb = 8}  &  \multicolumn{2}{c}{Emb = 12}  \\ \hline
    \quad Pre-train & 0.7877  & 0.4619 &  0.7944 & 0.4560 &  0.7962 & 0.4543 &  \textbf{0.7964} & \textbf{0.4541} \\ \hline
    \quad Fine-tune & 0.8016  & 0.4502 &  \textbf{0.8020} & \textbf{0.4491} &  0.8019 & 0.4494 &  0.8015 & 0.4496 \\ \Xhline{2\arrayrulewidth}
    Avazu & \multicolumn{2}{c|}{Emb = 4}   &  \multicolumn{2}{c|}{Emb = 8}  &  \multicolumn{2}{c|}{Emb = 16}  &  \multicolumn{2}{c}{Emb = 32}  \\ \hline
    \quad Pre-train & 0.7865 & 0.3816 & 0.7912 & 0.3732 &0.7916 & 0.3729 &\textbf{0.7916} & \textbf{0.3728}\\ \hline
    \quad Fine-tune & 0.7915 & 0.3732 & 0.7930 & 0.3722 & 0.7939 & 0.3719 & \textbf{0.7945} & \textbf{0.3713}\\    \Xhline{2\arrayrulewidth}
    XYZ & \multicolumn{2}{c|}{Emb = 2}   &  \multicolumn{2}{c|}{Emb = 4}  &  \multicolumn{2}{c|}{Emb = 8}  &  \multicolumn{2}{c}{Emb = 12}  \\ \hline
    \quad Pre-train & 0.8274 & 0.1375 & 0.8289  & 0.1366  & \textbf{0.8290} & \textbf{0.1365} & \textbf{0.8290} & \textbf{0.1365} \\ \hline
    \quad Fine-tune & 0.8307 & 0.1367 & \textbf{0.8366} &  \textbf{0.1354} & 0.8359 & 0.1358 & 0.8337  & 0.1361 \\  \toprule[1pt] 
\end{tabular}
\end{table*}

\begin{table*}[t]
    \centering
    \small
    \caption{Results of different MLP hidden sizes.}
    \label{tab:ablation_different_mlp_dimension}
    \begin{tabular}{l|cc|cc|cc|cc}
    \bottomrule[1pt]
    Datasets & AUC  & LogLoss & AUC  & LogLoss & AUC  & LogLoss & AUC  & LogLoss \\ \Xhline{2\arrayrulewidth}
    Criteo & \multicolumn{2}{c|}{MLP = 32}   &  \multicolumn{2}{c|}{MLP = 64}  &  \multicolumn{2}{c|}{MLP = 96}  &  \multicolumn{2}{c}{MLP = 128}  \\ \hline
    \quad Pre-train & \textbf{0.7964} & \textbf{0.4541} & \textbf{0.7964} & \textbf{0.4541} & \textbf{0.7964} & \textbf{0.4541} & \textbf{0.7964} & \textbf{0.4541} \\ \hline
    \quad Fine-tune & \textbf{0.8019} & \textbf{0.4496} & 0.8016 & 0.4498& 0.8018 & 0.4498 & 0.8016 & 0.4507  \\ \Xhline{2\arrayrulewidth}
    Avazu & \multicolumn{2}{c|}{MLP = 160}   &  \multicolumn{2}{c|}{MLP = 320}  &  \multicolumn{2}{c|}{MLP = 480}  &  \multicolumn{2}{c}{MLP = 640}  \\ \hline
    \quad Pre-train & \textbf{0.7913} & \textbf{0.3750} & \textbf{0.7913} & \textbf{0.3750} & \textbf{0.7913} & 0.3751 & \textbf{0.7913} & 0.3752\\ \hline
    \quad Fine-tune & 0.7961  & 0.3706 & 0.7964 & 0.3699& \textbf{0.7966} & \textbf{0.3697} &0.7964 &0.3698 \\   \Xhline{2\arrayrulewidth}
    XYZ & \multicolumn{2}{c|}{MLP = 32}   &  \multicolumn{2}{c|}{MLP = 64}  &  \multicolumn{2}{c|}{MLP = 96}  &  \multicolumn{2}{c}{MLP = 128}  \\ \hline
    \quad Pre-train & \textbf{0.8290}  & \textbf{0.1364} & \textbf{0.8290} & 0.1365 & \textbf{0.8290} & 0.1365 &  \textbf{0.8290} & 0.1365\\ \hline
    \quad Fine-tune & 0.8303  & 0.1371 &0.8306 & 0.1381 & \textbf{0.8334} & \textbf{0.1363} & 0.8323 & 0.1363 \\   \toprule[1pt] 
\end{tabular}
\end{table*}

We also explore the influence of different MLP hidden sizes. On the Criteo and XYZ AppGallery datasets, the original hidden layer dimension is 400, and here we low-rank decompose it into $\{32, 64, 96, 128\}$, respectively. On the Avazu dataset, the hidden layer dimension is 2000, and we compress it to $\{160, 320, 480, 640\}$. The results are shown in Table~\ref{tab:ablation_different_mlp_dimension}. Surprisingly, no matter what the size of the hidden layer is, the AUC is always the same when directly compressing. This indicates that the representation dimensionality required by the MLP layers of the CTR prediction models may be \emph{very low}. After further fine-tuning, compressed models with different MLP sizes obtained similar AUC, too.

In summary, we find that compared with  MLP sizes, the embedding dimensionalities have a higher influence on the models' performances.  Note that on the Avazu dataset, when the embedding dimension is 4, the compression ratio is 10.2$\times$, but it will slightly reduce the AUC. Here our goal is to reduce the dimensions as much as possible to achieve faster speeds, and we also want to ensure that the AUC and Logloss of the compressed model are comparable or even better than the original model. Therefore, we choose the present compression dimensions for DeepFM.  For simplicity and uniformity, we generalize DeepFM's hyperparameters to other CTR prediction models.

\subsubsection{Combine Our Framework with Mutil-Epoch Training }
In our original setting, we only train the baseline model 1 epoch because of the one-epoch phenomenon~\cite{liu2023multi}. To show that our method is also applicable to multi-epoch training, we imitate MEDA~\cite{liu2023multi} and reinitialize the embedding layer in each epoch and only keep other parts of the checkpoint (such as MLP layers). In this way, we can also achieve multi-epoch training. We train DeepFM with the XYZ AppGallery dataset. All hyperparameters are the same as in the original settings. The results are shown in Table~\ref*{tab:DeepFM_Mutil_Epoch_Training}.

\begin{table}
    \centering
    \small
    \caption{Results of multi-epoch training with DeepFM on the XYZ AppGallery dataset.}
    \label{tab:DeepFM_Mutil_Epoch_Training}
    \begin{tabular}{l|ccc}
    \bottomrule[1pt]
     Epoch & AUC &  LogLoss \\  \Xhline{2\arrayrulewidth}
     1   & 0.8290 &  0.1364 \\    
     2  &0.8318 & 0.1362 \\    
     3   &0.8339& 	0.1360 \\   
     4  &0.8325&0.1363\\   
     5&0.8311&0.1371\\  \hline
     Ours&\textbf{0.8362}&\textbf{0.1356}\\
     \toprule[1pt] 
    \end{tabular}
\end{table}

Therefore, even the multi-epoch training is still inferior to our framework when converging. Note that our framework is compressing the model, which requires less training time than multi-epoch training and consumes fewer resources during inference. For deep learning tasks, smaller model parameter sizes and computational overhead mean that the overall system requires fewer storage devices, computing units, and carbon emissions, especially for CTR models, which are widely used in the industry and inference millions of times a day, and bring direct economic benefits. 

Furthermore, our approach does not conflict with multi-epoch training, and we can also continue to perform our compression framework on multi-epoch trained models. Here we continue to compress the model based on 3-epoch training. The results are shown in Table~\ref{tab:appendix_DeepFM_Mutil_Epoch_Training_performing_ours}.

\begin{table}
    \centering
    \small
    \caption{Results of performing our framework after multi-epoch training with DeepFM on the XYZ AppGallery dataset.}
    \label{tab:appendix_DeepFM_Mutil_Epoch_Training_performing_ours}
    \begin{tabular}{l|ccc}
    \bottomrule[1pt]
     Method & AUC &  LogLoss \\  \Xhline{2\arrayrulewidth}
     Multi-epoch   & 0.8339&  0.1360 \\    
     ~~+ Ours  &0.8374&0.1350 \\    
     \toprule[1pt] 
    \end{tabular}
\end{table}

As the results show, when we use some additional techniques to multi-epoch train the model, we can continue to use our framework to further compress the model and achieve better results. This is what multi-epoch training and other state-of-the-art compression methods cannot achieve.

\subsubsection{One Possible Reason why the Compressed Model Outperforms the Original Model. }
It is an exceptional phenomenon that the compression model consistently outperforms the original model, and here we try to give a possible explanation.

In CTR fields, although the output features' dimensionality is large, there is a high linear correlation among them. Here is a concrete example. We collect the output features of three MLP layers on a pre-trained DeepFM with the Criteo dataset, and calculate the singular values of the output features. We report how many top-k singular values are needed when keeping 90\%, 95\%, and 99\% of the sum of total singular values.  The results are shown in Table~\ref{tab:appendix_top_k_singular}.

 \begin{table}[t]
    \centering
    \small
    \caption{The number of top-k singular values needed when keeping 90\%, 95\%, and 99\% of the sum of total singular values on the Criteo dataset with DeepFM.}
    \label{tab:appendix_top_k_singular}
    \begin{tabular}{l|ccc}
    \bottomrule[1pt]
     MLP Layer & 90\% &95\% &99\%\\ \hline
     1 &109 &139 &187 \\ 
    2 & 2 & 10 & 125	 \\ 
    3 &1& 6& 86	\\  \toprule[1pt] 
    \end{tabular}
\end{table}

Note that the feature dimensionality is 400. It shows in the MLP layers, only a few dimensions are needed to achieve 90\% singular values, and in the second and third FC, even only 1 or 2 dimensions are needed. So the output feature space is highly redundant and can be represented by fewer dimensions without loss of useful information. Discarding redundant principal components (by our framework) can help identify and extract the most important features, thereby improving processing efficiency and keeping the model accuracy. Besides, in CTR prediction tasks, models are prone to overfitting, hence they are often trained by one-epoch. With little or no decrease in AUC, our compression framework reduces dimensions with fewer principal components to remove unimportant redundant information, which is more like a regularization term to prevent the model from overfitting. Therefore, with our framework, the model can find better initial points and even better results can be obtained after further fine-tuning. Similar results can be found in previous papers, such as UMEC~\cite{shen2021umec}, which achieves a 50\% compression ratio and slightly increased AUC after further finetuning.


\subsection{Online Experiments}
\begin{table}[t]
    \centering
    \small
    \caption{Online experiment results for 7 days in comparison with non-compressed baseline.}
    \label{tab:online_test}
    \begin{tabular}{lcccc}
    \bottomrule[1pt]
    Method & AUC  & $\text{TP}_{\text{Avg}}$ & $\text{TP}_{\text{P99}}$ & AV \\ \hline
    Ours & +0.079\textperthousand & +15\% & +23\% & +2.34\% \\
    \toprule[1pt] & 
    \end{tabular}
\end{table}

To enhance the assessment of our method's efficacy, we integrated it into the online advertising system and executed a comprehensive 7-day online A/B test, aiming at contrasting our approach with the established baseline. We take the average throughput ($\text{TP}_{\text{Avg}}$) and the 99th percentile of throughput ($\text{TP}_{\text{P99}}$, which is usually used to assess the performance of a system under high load) to measure the inference efficiency. In addition to AUC, the Advertiser Value (AV) is selected to quantify the efficacy of the advertising investments. 

As illustrated in Table~\ref{tab:online_test}, similar to the observation in offline experiments, our method consistently improves both the AUC and throughput metrics. Note that compared to the baseline models, our method only improves online AUC by 0.079\textperthousand. This is because first, the original model's AUC is already very high, and second, the online experiment contains a large user base. The two difficulties make it difficult for the AUC to increase even slightly. However, our method also can achieve higher online AUC with faster speed. In addition, there is a substantial increase in Advertiser Value, marked by a significant boost of 2.34\%. This highlights that even when confronted with the complexities of online environments, our method maintains its efficiency and is proved to be highly effective.

\section{Conclusion}

In this paper, we believe that traditional low-rank decomposition methods tend to focus too much on model weights and ignore the distribution of feature maps. In contrast, we proposed a novel and unified low-rank decomposition framework for compressing CTR prediction models. Our framework mimics feature distribution and can be used in both the embedding tables and MLP layers of the CTR prediction models. Extensive experiments confirm that our framework can significantly increase the CTR prediction models' AUC while effectively reducing the parameter sizes and improving the throughput.

We find that since the inference bottleneck of CTR prediction models is mainly memory access, our framework cannot bring significant acceleration on GPU. Therefore, applying model compression in a reasonable way to accelerate the model's inference speed on GPU is an interesting future direction. Furthermore, our method can theoretically be applied to a wide variety of recommendation models, so we will continue to extend our method to these models in the future.

\section{Acknowledgments}
We acknowledge the funding provided by the National Natural Science Foundation of China under Grant 62276123 and Grant 61921006. J. Wu is the corresponding author.

\bibliographystyle{ACM-Reference-Format}
\bibliography{references}

\appendix

\newpage

\section{Appendix}


\subsection{Illustration of the Compression Process}
\begin{figure}[t]
  \centering
 \includegraphics[width=0.98\linewidth]{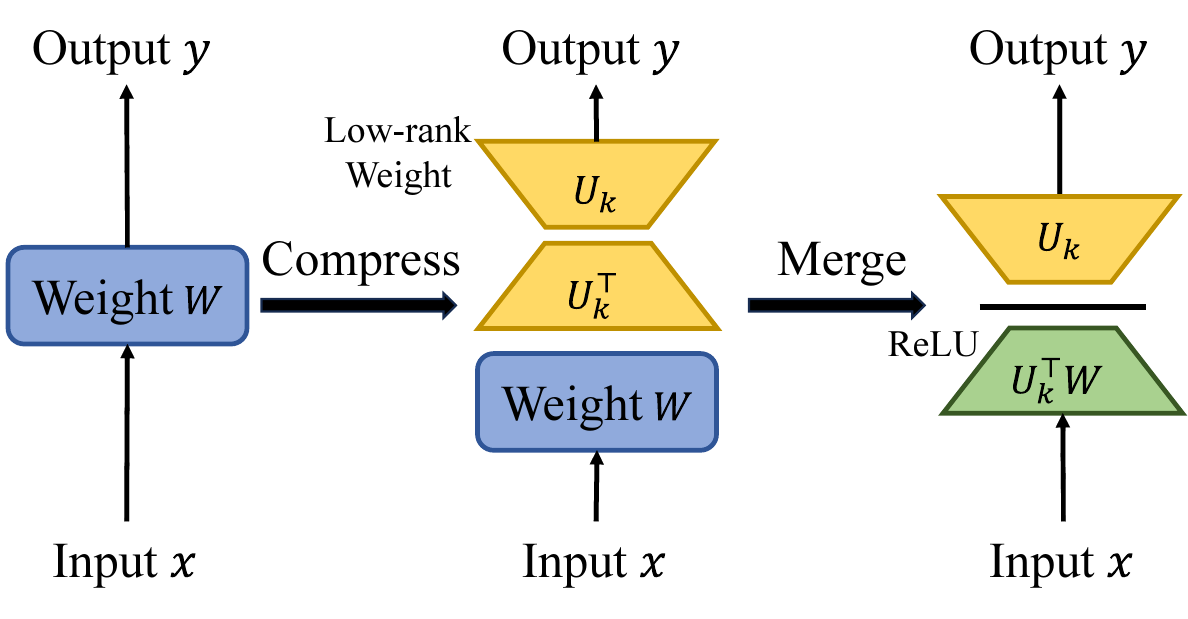}
  \caption{Illustration of compressing MLP layers.}
  \Description{Illustration of compressing MLP layers.}
  \label{fig:compress_fc} 
\end{figure}

\begin{figure}[t]
  \centering
 \includegraphics[width=0.98\linewidth]{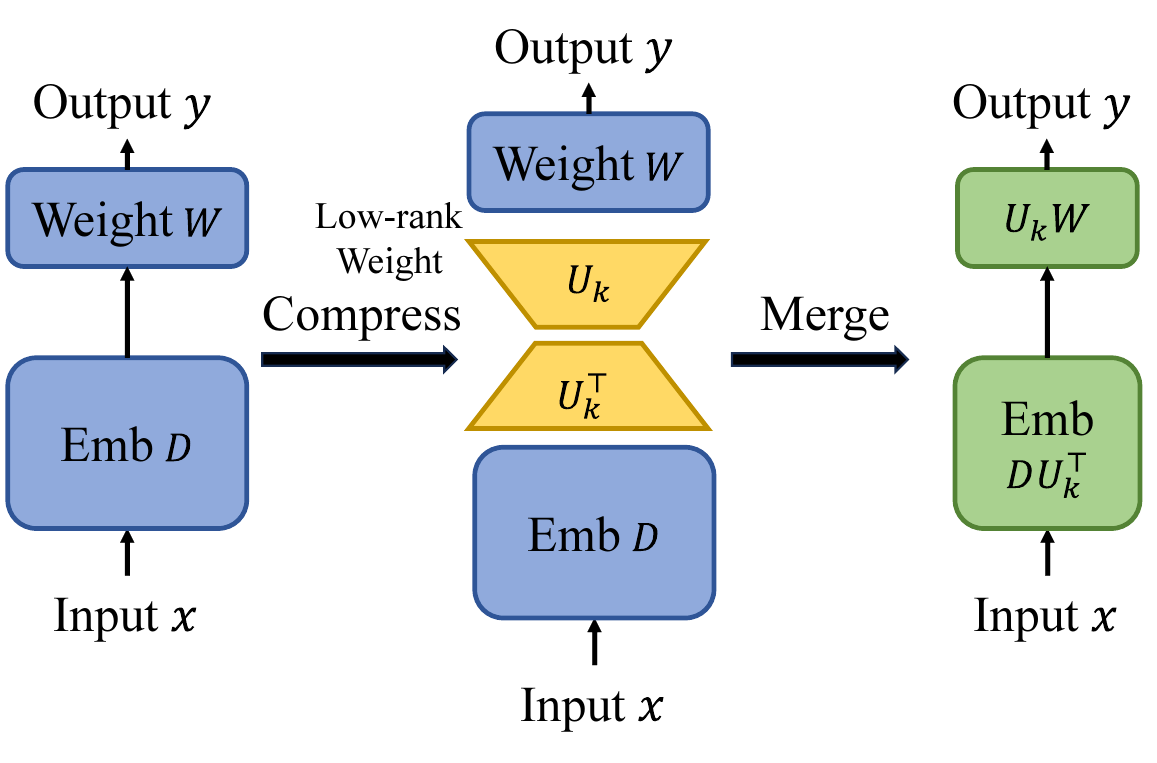}
  \caption{Illustration of compressing embedding tables.}
  \Description{Illustration of compressing embedding tables.}
  \label{fig:compress_emb} 
\end{figure}

We show the process of compressing the MLP layer and embedding tables in Figures~\ref{fig:compress_fc} and~\ref{fig:compress_emb}, respectively. The figures include the calculation of low-rank weights and fusion methods. It is worth noting that when compressing embedding tables, if there is an additional interaction layer between embedding and the first FC weight, then the fusion of $W$ and $U_k$ is not applicable. 

\subsection{Detailed Training Settings}\label{sec:TrainingSettings}

When training the baseline model, we follow the training settings in FiBiNet, i.e., we use Adam with a mini-batch size of 2000, 1000, and 500 for the XYZ AppGallery, Criteo, and Avazu datasets, respectively. The learning rate is set to 0.0001 and we also apply dropout and set the rate to 0.5.

When fine-tuning the MLP-compressed models, the learning rate is 0.001. On the Criteo and XYZ AppGallery datasets, we set the batch size to 20000, and on Avazu the batch size is 10000. When further fine-tuning the compressed models with reduced embedding dimensions, we set the batch size to 10000, 5000, and 3000 on the Criteo, Avazu, and XYZ AppGallery datasets, respectively. We also set the learning rate to 0.001 and remove dropout on the Criteo and Avazu datasets. On the XYZ AppGallery datasets, we set the dropout rate as 0.3. In all experiments, the weight decay is 1e-3, and we sum the binary cross-entropy loss and the l2-norm of model weights as loss targets, i.e.,
\begin{equation}
\text { L }=-\frac{1}{N} \sum_{i=1}^N\left(y_i \log \left(\hat{y}_i\right)+\left(1-y_i\right) \log \left(1-\hat{y}_i\right)\right) + r \sum_{i=1}^M \|W_i \|^2,
\end{equation}
In the first item of the right hand side, $y_i$ is the ground truth of the $i$-th instance, $\hat{y}_i$ is the model's prediction, and $N$ is the total size of samples. Correspondingly, in the second item, $r$ is the loss ratio, $W_i$ is the $i$-th model weight and $M$ is the number of model weights. When compressing the embedding dimension on the XYZ AppGallery dataset, we set $r$ to 1e-2, and in other cases, the ratio $r$ is 1e-5. In all experiments, we set the random seed as 0.

It is worth noting that we did not mention validation dataset. This is because we find that the model consistently behaves the same on the training and test set when there are different training hyperparameters. Taking DeepFM with AFM MLP on the Criteo dataset as an example, we set the learning rate in \{1e-2,5e-3,1e-3,5e-4,1e-4\}. The results are shown in Table~\ref{tab:appendix_DeepFM_lr_AFM_MLP}.

As we can see from the table, it is enough that we only need to adjust the hyperparameters so that the model performs well on the training set. 

Besides, to prove the practicality and ease-to-use of our method, we do \emph{not} deliberately tune the hyperparameters of the model training and simply generalize DeepFM's training hyperparameters to other models.

\subsection{Training Times of the Baseline and Our Compression Framework}

To prove the effectiveness of our compression framework, we also report the GPU times of the baseline model and our compression framework during the training process. Note that AFM for MLP and embedding tables processes only infer the training dataset once and then directly compute the compression weights. The inference computational resources it requires are small and the time it takes is negligible. As for the further fine-tuning process, our framework only need to fine-tune the sub-model 1 epoch after each compression process, so the training time required for the whole framework is short. Taking DeepFM as an example, we train the model on a single 3090 GPU on the Criteo, Avazu, and XYZ datasets. The results are shown in Table~\ref*{tab:appendix_DeepFM_Training_Times}.

\begin{table}
    \centering
    \small
    \caption{DeepFM's performances of different learning rate on the Criteo dataset with AFM MLP.}
    \label{tab:appendix_DeepFM_lr_AFM_MLP}
    \begin{tabular}{l|ccccc}
    \bottomrule[1pt]
     Learning rate                      & 1e-2&5e-3&1e-3&5e-4&1e-4  \\  \Xhline{2\arrayrulewidth}
     Train AUC   & 0.8198 & 	0.8321 & 	\textbf{0.8396} & 	0.8375 & 	0.8278 \\    \hline
     Train LogLoss & 	0.4415 & 	0.4208 & 	\textbf{0.4153}	 & 0.4174 & 	0.4254 \\    \hline
     Test AUC  &0.7534  &	0.7786	  &\textbf{0.8016 } &	0.8008	  &0.7998 \\    \hline
    Test LogLoss  &0.5163  &0.4733  &\textbf{0.4498}  &0.4503  &	0.4515 \\    \hline
     \toprule[1pt] 
    \end{tabular}
\end{table}
\begin{table}
    \centering
    \small
    \caption{Results of training times with DeepFM.}
    \label{tab:appendix_DeepFM_Training_Times}
    \begin{tabular}{l|ccc}
    \bottomrule[1pt]
     Dataset                      & Criteo &	Avazu&	XYZ  \\  \Xhline{2\arrayrulewidth}
     Original Model   & 3h39m04s	& 5h57m40s  & 20m10s \\    \hline
     AFM MLP  & 9m47s	& 16m47s & 1m43s \\    \hline
     AFM EMB   & 8m6s	& 14m40s & 7m21s \\    \hline
     \toprule[1pt] 
    \end{tabular}
\end{table}

Note that the training time is proportional to the batch size. It can be seen that the training time our framework needs is very short, and a compression time of tens of minutes is acceptable. It takes longer to train AFM EMB on the XYZ AppGallery dataset than AFM MLP, because we set batch sizes as 20000 when compressing the MLP layer and batch sizes as 3000 when reducing the embedding dimensions.


\subsection{Compare Our Framework with Other Compression Techniques }

To demonstrate the superiority of our approach, we train DeepFM with Criteo and compare our framework with other three state-of-the-art methods, i.e., UMEC (pruning method)~\cite{shen2021umec}, Product Quantization (PQ, quantization method)~\cite{Herve2011Product} and QR (hashing method)~\cite{Shi2020Compositional}. The results are shown in Table~\ref{tab:appendix_afm_comprare_other_sota}. It can be seen that previous works hardly achieve AUC improvement with a 3-5x compression ratio and faster inference.  But, when AUC decreases even 0.1\%, it will cause revenue loss and will not be adopted. Our method can achieve both higher AUC and faster speed in compressing CTR models.
 \begin{table}[t]
    \centering
    \small
    \caption{Comparison of our framework and other compression techniques on the Criteo dataset with DeepFM.}
    \label{tab:appendix_afm_comprare_other_sota}
    \begin{tabular}{l|ccc}
    \bottomrule[1pt]
     Dataset                      & Param. (M) & AUC & LogLoss\\ \hline
     Baseline &   574.46 & 0.7964 & 0.4541 \\  \Xhline{2\arrayrulewidth}
    UMEC & 131.09	&0.7960	&0.4550	 \\ 
    PQ &102.08 &0.7929	&0.4578	\\ 
    QR &106.38 &0.7937	&0.4569	\\ 
    Ours &\textbf{101.42} & \textbf{0.8021} &\textbf{0.4495}	  \\ \toprule[1pt] 
    \end{tabular}
\end{table}

\subsection{Transfer the Compression Weight into Different Datasets.}
Here we use MLP compression weights calculated on the XYZ dataset to compress DeepFM trained on Criteo, and then fine-tune the compressed model with Criteo. All hyperparameters are the same as in the previous experiments.
 \begin{table}[t]
    \centering
    \small
    \caption{Transfer the XYZ compression weight into Criteo dataset with DeepFM.}
    \label{tab:appendix_transfer_compression_weight}
    \begin{tabular}{l|ccc}
    \bottomrule[1pt]
     - & AUC &	LogLoss\\ \hline
     Pre-train&	0.7404&	0.7306 \\ 
    Fine-tune&	0.8009&	0.4506\\  \toprule[1pt] 
    \end{tabular}
\end{table}

These results in Table~\ref{tab:appendix_transfer_compression_weight} show that transferring the weight from another dataset causes a large drop in AUC, but AUC also recovers after fine-tuning. Although this result is not as good as using the original dataset directly (0.8016 v.s. 0.8009), it is better than the original model (0.7964).

\subsection{Model Throughput on the GPU Device}

In this section, we report the throughput of the original model and various compression models on the GPU device. Here we take DeepFM on GTX 3090 GPU as examples. Note that because SVD and AFM share the same model structure, here we ignore SVD and report the throughput of the AFM and TTD compression models. The mini-batch size is still 10000.
    
\begin{table}[t]
    \centering
    \small
    \caption{Parameter sizes and throughput on GPU of AFM and TTD on the Criteo, Avazu, and XYZ AppGallery datasets with DeepFM.}
    \label{tab:appendix_afm_ttd_throughput_gpu}
    \begin{tabular}{l|cc}
    \bottomrule[1pt]
     Dataset                      & Param. (M) & Throughput (sample/s)\\ \hline
     {Criteo} &   574.46 &  2675251.43 \\  \Xhline{2\arrayrulewidth}
     \quad+AFM MLP   &  574.24  &  \textbf{2721113.99} (+1.71\%) \\   \hline
     \quad+AFM EMB   &    101.42    &  2602868.49 (-2.70\% ) \\   \Xhline{2\arrayrulewidth}
     \quad+TTD EMB   & 3.76    & 2110.71  (-99.92\%) \\   \Xhline{2\arrayrulewidth}
   {Avazu} &   492.13 &   \textbf{4098317.76}  \\  \Xhline{2\arrayrulewidth}
     \quad+AFM MLP   &  486.69  & 3958743.08 (-3.41\% )   \\   \hline
     \quad+AFM EMB   &    87.98    &  3708437.38  (-9.51\% )  \\   \Xhline{2\arrayrulewidth}
     \quad+TTD EMB   &  12.21    &  2602.00   (-99.93\% ) \\    \Xhline{2\arrayrulewidth}
     {XYZ} &   633.61 &  \textbf{1398675.31} \\  \Xhline{2\arrayrulewidth}
     \quad+AFM MLP   &  633.39  &   1335091.61  (-4.55\% )\\   \hline
     \quad+AFM EMB   &    186.35    & 1285443.42 (-8.10\% )\\   \Xhline{2\arrayrulewidth}
     \quad+TTD EMB   &  12.00    &  1897.94  (-99.86\%)\\    \toprule[1pt] 
    \end{tabular}
\end{table}

The results are shown in Table~\ref{tab:appendix_afm_ttd_throughput_gpu}. During the throughput testing process, we find that the GPU utilization is less than 20\%, which is very low. This phenomenon shows that the throughput on the GPU cannot truly reflect the effectiveness of our algorithms. However, we still achieve a 3-5x reduction in memory usage and comparable throughput. In contrast, TTD needs to recalculate the embedding layer on each field before it can obtain the corresponding feature embedding. This operation seriously affects its speed. These results indicate that our framework is far superior to traditional tensor decomposition methods, both on the GPU and CPU devices.

\begin{table*}[t]
    \centering
    \small
    \caption{DeepFM's performances on the Criteo and Avazu datasets with different random seeds.}
    \label{tab:appendix_DeepFM_different_seed}
    \begin{tabular}{ll|cc||cc}
    \bottomrule[1pt]
    \multicolumn{2}{l|}{DeepFM}                       &AUC  & Logloss & AUC  & Logloss \\ \hline
    \multicolumn{2}{l|}{Baseline} & 0.7966 $\pm$ 0.0002 &	0.4543  $\pm$ 0.0002	&0.7919  $\pm$ 0.0002	&0.3727  $\pm$ 0.0003  \\  \Xhline{2\arrayrulewidth}
    \multicolumn{1}{l|}{\multirow{2}{*}{~~+AFM MLP}} & Pre-train &  0.7965 $\pm$ 0.0002	 &0.4536 $\pm$ 0.0002	&0.7902 $\pm$ 0.0008	&0.3740 $\pm$ 0.0018\\ \cline{2-6}  
    \multicolumn{1}{l|}{}                         & Fine-tune & 0.8014  $\pm$ 0.0003&	0.4500 $\pm$ 0.0003	&\textbf{0.7969 $\pm$ 0.0002}	&\textbf{0.3700  $\pm$ 0.0002}\\ \hline
    \multicolumn{1}{l|}{\multirow{2}{*}{~~+AFM EMB}} & Pre-train  & 0.7901 $\pm$ 0.0002	 & 0.4610 $\pm$ 0.0007	 & 0.7940 $\pm$ 0.0003 & 	0.3721 $\pm$ 0.0003\\ \cline{2-6}  
    \multicolumn{1}{l|}{}                         & Fine-tune  & \textbf{0.8022 $\pm$ 0.0002} & \textbf{0.4493 $\pm$ 0.0003} &	0.7950 $\pm$ 0.0001	 & 0.3718 $\pm$ 0.0002 \\       \toprule[1pt] 
    \end{tabular}
\end{table*}

\begin{table*}[t]
    \centering
    \small
    \caption{Results of AFM on the Criteo (columns 3--4), Avazu (columns 5--6), and XYZ AppGallery (columns 7--8) datasets with each model.}
    \label{tab:appendix_afm_each_model}
    \begin{tabular}{ll||cc||cc||cc}
    \bottomrule[1pt]
    \multicolumn{2}{l||}{Model}                       &AUC  & Logloss &AUC  & Logloss &AUC  & Logloss \\ \hline
    \multicolumn{2}{l||}{DCN} &0.7932 & 0.4570 & 0.7890 & 0.3745  & 0.8271 & 0.1373\\  \Xhline{2\arrayrulewidth}
    \multicolumn{1}{l|}{\multirow{2}{*}{~~+AFM MLP}} & Pre-train  & 0.7925 & 0.4574 & 0.7880 & 0.3778  & 0.8271 & 0.1374 \\ \cline{2-8} 
    \multicolumn{1}{l|}{}                         & Fine-tune &0.8013  & 0.4496  & 0.7933 & 0.3719  & 0.8299 & 0.1378 \\ \hline
    \multicolumn{1}{l|}{\multirow{2}{*}{~~+AFM EMB}} & Pre-train  & 0.7949 & 0.4580  & 0.7930 & 0.3720 & 0.8299 & 0.1378  \\ \cline{2-8} 
    \multicolumn{1}{l|}{}                         & Fine-tune & \textbf{0.8023} & \textbf{0.4489}  &  \textbf{0.7941} &  \textbf{0.3715} & \textbf{0.8310} & \textbf{0.1363} \\ \Xhline{2\arrayrulewidth}
    \multicolumn{2}{l||}{NFM} & 0.7929     & 0.4571 & 0.7864  & 0.3764  &0.8330 & 0.1370\\  \Xhline{2\arrayrulewidth}
    \multicolumn{1}{l|}{\multirow{2}{*}{~~+AFM MLP}} & Pre-train  & 0.7900 & 0.4650  &0.7862& 0.3906 & 0.8329 & 0.1371  \\ \cline{2-8} 
    \multicolumn{1}{l|}{}                         & Fine-tune  &   0.8016 & 0.4492  &  \textbf{0.7917} & \textbf{0.3729}  &0.8300 & 0.1369 \\ \hline
    \multicolumn{1}{l|}{\multirow{2}{*}{~~+AFM EMB}} & Pre-train  & 0.7941 & 0.4573 & 0.7856 & 0.3770  & 0.8300 & 0.1369\\ \cline{2-8} 
    \multicolumn{1}{l|}{}                         & Fine-tune  &  \textbf{0.8025} & \textbf{0.4484} &  0.7911 &  0.3738 & \textbf{0.8336} &	\textbf{0.1359} \\ \Xhline{2\arrayrulewidth}
    \multicolumn{2}{l||}{AutoInt} &0.7939 & 0.4563  & 0.7904 &  0.3735 & 0.8266 & 0.1371 \\  \Xhline{2\arrayrulewidth}
    \multicolumn{1}{l|}{\multirow{2}{*}{~~+AFM MLP}} & Pre-train  & 0.7935  & 0.4566   & 0.7902 & 0.3738  & 0.8269 & 0.1371 \\ \cline{2-8} 
    \multicolumn{1}{l|}{}                         & Fine-tune  & 0.8016 & 0.4496 & \textbf{0.7957}  & \textbf{0.3705}  & \textbf{0.8348} & \textbf{0.1358} \\ \hline
    \multicolumn{1}{l|}{\multirow{2}{*}{~~+AFM EMB}} & Pre-train  & 0.7938 & 0.4583   & 0.7948 & 0.3709  & \textbf{0.8348} & \textbf{0.1358} \\ \cline{2-8} 
    \multicolumn{1}{l|}{}                         & Fine-tune  &\textbf{0.8019} & \textbf{0.4494}  &  0.7952 & 0.3708  & 0.8312 &	0.1365\\ \Xhline{2\arrayrulewidth}
    \multicolumn{2}{l||}{FiBiNet}  &0.8002 & 0.4509 & 0.7965 & 0.3699  &0.8341 & \textbf{0.1350} \\  \Xhline{2\arrayrulewidth}
    \multicolumn{1}{l|}{\multirow{2}{*}{~~+AFM MLP}} & Pre-train  & 0.7761 & 0.4994   & 0.7951 & 0.3754  & 0.8341 & \textbf{0.1350} \\ \cline{2-8} 
    \multicolumn{1}{l|}{}                         & Fine-tune  & \textbf{0.8055}  & \textbf{0.4458}  & \textbf{0.8011}  & \textbf{0.3668}  &0.8326 & 0.1363  \\ \hline
    \multicolumn{1}{l|}{\multirow{2}{*}{~~+AFM EMB}} & Pre-train  & 0.7954 & 0.4549   & 0.7919 & 0.3737  & 0.8329 & 0.1363 \\ \cline{2-8} 
    \multicolumn{1}{l|}{}                         & Fine-tune  & 0.8051 & 0.4462 & 0.7968 & 0.3699 & \textbf{0.8372} &	0.1356 \\   \Xhline{2\arrayrulewidth}

    \multicolumn{2}{l||}{DCNv2}  &0.7947 & 0.4562 & 0.7913 & 0.3732  & 0.8339& 0.1350  \\  \Xhline{2\arrayrulewidth}
    \multicolumn{1}{l|}{\multirow{2}{*}{~~+AFM MLP}} & Pre-train  & 0.7948 & 0.4565   & 0.7912 & 0.3731& 0.8339& 0.1350 \\ \cline{2-8} 
    \multicolumn{1}{l|}{}                         & Fine-tune  & 0.8026  & 0.4486  & \textbf{0.7959}  & \textbf{0.3704}  & 0.8343& 0.1349  \\ \hline
    \multicolumn{1}{l|}{\multirow{2}{*}{~~+AFM EMB}} & Pre-train  & 0.7964 & 0.4545   & 0.7943 & 0.3719  & 0.8342& 0.1349\\ \cline{2-8} 
    \multicolumn{1}{l|}{}                         & Fine-tune  & \textbf{0.8029} & \textbf{0.4485} & 0.7950 & 0.3709 & \textbf{0.8347}& \textbf{0.1344} \\   \Xhline{2\arrayrulewidth}

    \multicolumn{2}{l||}{GDCN}  &0.7949 & 0.4559 & 0.7913 & 0.3729  &0.8315& 0.1358  \\  \Xhline{2\arrayrulewidth}
    \multicolumn{1}{l|}{\multirow{2}{*}{~~+AFM MLP}} & Pre-train  & 0.7948 & 0.4558  & 0.7913 & 0.3730& 0.8315& 0.1358 \\ \cline{2-8} 
    \multicolumn{1}{l|}{}                         & Fine-tune  & 0.8015 & 0.4496  & \textbf{0.7962}  & \textbf{0.3701}  & \textbf{0.8397}& \textbf{0.1347}  \\ \hline
    \multicolumn{1}{l|}{\multirow{2}{*}{~~+AFM EMB}} & Pre-train  & 0.7961 & 0.4551   & 0.7945 & 0.3713 &  0.8396& \textbf{0.1347}\\ \cline{2-8} 
    \multicolumn{1}{l|}{}                         & Fine-tune  & \textbf{0.8022} & \textbf{0.4491} & 0.7953 & 0.3706  &  0.8348& 0.1361 \\   

    \toprule[1pt] 
    \end{tabular}
\end{table*}

\subsection{The Influence of Different Random Seeds}
To prove that our method has a statistical improvement, we use 5 different random seeds to divide the Criteo and Avazu datasets and compress the DeepFM model (Note that XYZ training and test datasets are already partitioned and cannot be changed). We report the AUC and LogLoss's mean and standard deviation of the original compressed models.

The results are shown in Table~\ref{tab:appendix_DeepFM_different_seed}. With different random seeds, our method has always improved steadily. This shows that our framework is statistically significant enough to report an improvement in accuracy.

\subsection{The Detailed Performances of AFM in Each Model}
In this subsection, we present detailed performances of direct compression and further fine-tuning on the other six DLRMs. Similarly, the results of AFM on the Criteo, Avazu, and XYZ AppGallery datasets are shown in Table~\ref{tab:appendix_afm_each_model}. Since we have already reported parameter sizes and throughput in the main text, we ignore these results here.

We can draw a similar conclusion to the above text from those results, i.e., when compressing the MLP layers and embedding tables, our algorithm will only bring a slight drop in AUC, or even no decrease in it. After further fine-tuning, the AUC of the model will further increase. It is worth noting that on the XYZ AppGallery dataset, when we perform AFM directly on AutoInt, FiBiNet and GDCN, sometimes the model's AUC even increases, but after further fine-tuning, the AUC also may decrease. This phenomenon shows that in some cases, directly applying AFM to compress model can already achieve good performances, and the sub-models no longer need fine-tuning.

\subsection{Compare Our Framework with Train from Scratch }

Our framework first trains a large network first and then compress it into a small network. We now compare our ``train-compress-finetune'' framework and directly ``train from scratch'' (refer to TFS) a small network. The results are shown in Table~\ref*{tab:appendix_tfs}. 

\begin{table*}[t]
    \centering
    \small
    \caption{Comparison of our framework and training from scratch.}
    \label{tab:appendix_tfs}
    \begin{tabular}{ll|ll|ll|ll}
    \bottomrule[1pt]
    \multicolumn{2}{l|}{DeepFM} & Criteo & &	Avazu&&	XYZ&  \\  \Xhline{2\arrayrulewidth}
    \multicolumn{2}{l|}{Strategy}  & 	TFS& 	Ours	& TFS	& Ours	& TFS& 	Ours \\ \hline
    \multicolumn{1}{l|}{\multirow{2}{*}{AFM MLP}} & AUC  &0.7997&\textbf{0.8016}&	0.7944&	\textbf{0.7964}	&0.8304&	\textbf{0.8306}\\    \cline{2-8}
    \multicolumn{1}{l|}{} & LogLoss  & 0.4511	 &\textbf{0.4498 }&	0.3712	 &\textbf{0.3699} &	0.1385	 &\textbf{0.1381}\\    \hline
    \multicolumn{1}{l|}{\multirow{2}{*}{AFM EMB}} & AUC  & 0.7961	&\textbf{0.8021}&	0.7888&	\textbf{0.7948}	&0.8360&	\textbf{0.8362}\\    \cline{2-8}
    \multicolumn{1}{l|}{} & LogLoss  & 	0.4545&\textbf{0.4495}	&0.3746&	\textbf{0.3716}&	0.1364&	\textbf{0.1356}\\    \hline
     \toprule[1pt] 
    \end{tabular}
\end{table*}
\begin{table*}[t]
    \centering
    \small
    \caption{Results of different compression orders.}
    \label{tab:ablation_compression_order}
    \begin{tabular}{ll|cc|ll|cc}
    \bottomrule[1pt]
    \multicolumn{2}{l|}{Criteo}                       &AUC  & Logloss &\multicolumn{2}{l|}{Avazu}  &AUC  & Logloss \\ \hline
    \multicolumn{1}{l|}{\multirow{2}{*}{\quad+AFM MLP}} & Pre-train  & 0.7964 & 0.4541  & \multicolumn{1}{l|}{\multirow{2}{*}{\quad+AFM MLP}}  & Pre-train  &  0.7913 & 0.3750\\ \cline{2-4}\cline{6-8}
    \multicolumn{1}{l|}{}                        & Fine-tune & 0.8016 & 0.4498  & \multicolumn{1}{l|}{}                        & Fine-tune & 0.7964  & 0.3699\\ \hline
    \multicolumn{1}{l|}{\multirow{2}{*}{\quad+AFM EMB}} & Pre-train  & 0.7900 & 0.4601 & \multicolumn{1}{l|}{\multirow{2}{*}{\quad+AFM EMB}}  & Pre-train &  0.7942 & 0.3720 \\ \cline{2-4}\cline{6-8}
    \multicolumn{1}{l|}{}                         & Fine-tune  &\textbf{0.8021} & \textbf{0.4493} & \multicolumn{1}{l|}{}                        & Fine-tune & 0.7948 & 0.3716  \\ \Xhline{2\arrayrulewidth}
    \multicolumn{1}{l|}{\multirow{2}{*}{\quad+AFM EMB}} & Pre-train  & 0.7877 & 0.4619  & \multicolumn{1}{l|}{\multirow{2}{*}{\quad+AFM EMB}}  & Pre-train & 0.7912 & 0.3732 \\ \cline{2-4}\cline{6-8}
    \multicolumn{1}{l|}{}                        & Fine-tune & 0.8000  & 0.4510  & \multicolumn{1}{l|}{}                        & Fine-tune & 0.7933 & 0.3721\\ \hline
    \multicolumn{1}{l|}{\multirow{2}{*}{\quad+AFM MLP}} & Pre-train  & 0.7955 & 0.4581 & \multicolumn{1}{l|}{\multirow{2}{*}{\quad+AFM MLP}}  & Pre-train & 0.7239 & 0.4239 \\ \cline{2-4}\cline{6-8}
    \multicolumn{1}{l|}{}                         & Fine-tune  & 0.8018 & 0.4494 & \multicolumn{1}{l|}{}                        & Fine-tune &  \textbf{0.7956} & \textbf{0.3706} \\    \toprule[1pt] 
    \end{tabular}
\end{table*}

The first two lines mean the AUC and LogLoss of only performing AFM for MLP layers, and last two lines mean the AUC and LogLoss of performing AFM both for MLP layers and embedding tables. We can see that our ``train-compress-finetune'' framework consistently outperforms the ``training from scratch'' approach.

\subsection{The Order of Compressing Embedding and MLP Layers.} 

In previous experiments, we first compress the MLP layers and then continue to reduce the embedding dimensions. In this subsection, we research the influence of the compression order.

Table~\ref{tab:ablation_compression_order} shows the influence of different compression orders on the DLRMs' performances. The first four lines represent that we compress MLP first and then reduce the embedding dimension. Correspondingly, the last four lines represent that we first compress embedding. It can be seen that the final AUC difference between the two strategies is not large, and each strategy does not show a sustained lead over the other strategy.

\end{document}